\documentclass[conference]{IEEEtran}

\usepackage[normalem]{ulem}
\usepackage{comment}
\usepackage{mathrsfs}
\usepackage{amsmath}

\usepackage{breqn}
\usepackage{microtype}
\usepackage{caption}
\usepackage{subcaption}
\usepackage{url}
\usepackage{hyperref}

\usepackage{soul}
\usepackage{booktabs} 
\usepackage{setspace}
\usepackage{setspace}
\usepackage{subcaption}
\usepackage{verbatim}
\usepackage{courier}
\usepackage{listings}
\usepackage{enumitem}
\usepackage{xspace}
\usepackage{enumitem}

\usepackage[colorinlistoftodos]{todonotes}
\usepackage{xcolor}

\usepackage{xr}
\definecolor{mygreen}{rgb}{0,0.6,0}
\usepackage[ruled,vlined]{algorithm2e}
\usepackage[normalem]{ulem}
\usepackage{multirow}
\usepackage{float}
\usepackage{graphicx}
\usepackage{caption}
\usepackage{subcaption}

\usepackage{enumitem}
\setitemize{noitemsep,topsep=0pt,parsep=0pt,partopsep=0pt}

\newcommand{\squeezeup}{\vspace{-1mm}}

\newcommand{\frameworkname}{Estimator}

\begin{document}
\title{MATCH: An MPI Fault Tolerance Benchmark Suite}

\author{\IEEEauthorblockN{Luanzheng Guo\IEEEauthorrefmark{2},
Giorgis Georgakoudis\IEEEauthorrefmark{3}, Konstantinos Parasyris\IEEEauthorrefmark{3},
Ignacio Laguna\IEEEauthorrefmark{3}
and
Dong Li\IEEEauthorrefmark{2}
}
\IEEEauthorblockA{
\IEEEauthorrefmark{2}
\textit{EECS}, University of California, Merced, USA; 
  \{lguo4, dli35\}@ucmerced.edu 
}
\IEEEauthorblockA{
  \IEEEauthorrefmark{3}
  \textit{CASC}, Lawrence Livermore National Laboratory, USA;
     \{georgakoudis1, parasyris1, lagunaperalt1\}@llnl.gov
  }
}

\maketitle
%
%
\begin{abstract}
MPI has been ubiquitously deployed in flagship HPC systems aiming to accelerate distributed scientific applications running on tens of hundreds of 
processes and compute nodes. Maintaining the correctness and integrity of MPI application execution is critical, especially for  
safety-critical scientific applications. Therefore, a collection of effective MPI fault tolerance techniques have been proposed to enable MPI 
application execution to efficiently resume from system failures. 
However, there is no structured way to study and compare different MPI fault tolerance designs, so to guide the selection and development of efficient 
MPI fault tolerance techniques for distinct scenarios. To solve this problem, we design, develop, and evaluate a benchmark suite called MATCH to characterize, 
research, and comprehensively compare different combinations and configurations of MPI fault tolerance designs. 
Our investigation derives useful findings: 
(1) Reinit recovery in general performs better than ULFM recovery; 
(2) Reinit recovery is independent of the scaling size and the input problem size, whereas ULFM recovery is not;
(3) Using Reinit recovery with FTI checkpointing is a highly efficient fault tolerance design.
MATCH code is available at \url{https://github.com/kakulo/MPI-FT-Bench}.
\end{abstract}
\section{Introduction}
\label{sec:intro}
As supercomputers continue to increase in computational power and size, next-generation HPC systems are
expected to incur a much higher failure rate than current systems. 
For example, the Sequoia supercomputer located in Lawrence Livermore National Laboratory (LLNL)
reported a mean time between node failures of 19.2 hours in 2013~\cite{dongarra2013emerging}. 
After that, in 2014, the Blue Waters supercomputer reported a mean time between node failures of 6.7 hours~\cite{di2014lessons}. 
Most recently, the Taurus system located in TU Dresden reported a mean time between node failures of 3.65 hours~\cite{ghi2016lessons}.

This trend raises concerns in the HPC community for MPI applications 
running on tens of thousands of processes and nodes, that are prone to fail due to the increased probability of a failure.
Process and node failures frequently occur in production HPC systems due to power outages and other issues. 
MPI process and node failures are usually fail-stop failures. In this type of failure, application execution cannot
continue without repairing the communication and has to stop. 

These crucial facts lead to increasing importance of and challenges for developing efficient and effective 
fault tolerance designs for scaling HPC systems~\cite{sc18:fliptracker,ipdps19:guo}. 
There are numerous fault tolerance techniques proposed to protect MPI application execution from system failures. 
MPI fault tolerance techniques can be assigned into two types of different focus. 
Checkpointing~\cite{katti2015scalable,herault2015practical,bouteiller2015plan,ali2016complex,Laguna:2014:EUF:2642769.2642775}, commonly used in HPC applications,
is one type of fault tolerance technique that focuses on restoring application state.
Checkpointing takes place in two separate phases: storing the system state and recovering from it 
in case of a failure.
Checkpointing helps MPI applications quickly restore application state from the latest checkpoints through saving application execution state periodically.
The other type of MPI fault tolerance technique focuses on restoring the MPI state.
Restarting is a baseline solution for restoring the MPI state, which immediately restarts an application 
after execution collapses due to a failure. Later, because of
the inefficiency of restarting an application, HPC practitioners 
propose MPI recovery mechanisms to restore the MPI state online. User-level Fault Mitigation (ULFM)~\cite{bland2013post} 
and Reinit~\cite{laguna2016evaluating,doi:10.1002/cpe.4863,isc2020reinit} are the two pioneering MPI recovery frameworks in this effort.
ULFM supports a wide range of recovery strategies, including local forward recovery and global-restart recovery, whereas Reinit only supports global-restart recovery. 
ULFM is a powerful MPI recovery framework but complicated to use. In contrast, Reinit requires less programming effort. 

However, there is no existing framework that enables a comprehensive comparison between different MPI fault tolerance techniques. 
To solve the problem, we design and develop a benchmark suite called MATCH, aiming to study the 
performance efficiency of different MPI fault tolerance configurations.
MATCH contains six proxy applications from 
the Exascale Computing Project (ECP) Proxy Apps Suite and LLNL Advanced Simulation and Computing (ASC) proxy application suite; MATCH uses Fault Tolerance Interface (FTI)~\cite{6114441} for data recovery 
and uses ULFM and Reinit for MPI recovery. 
We pick a representative set of HPC applications, but our methodology is extensible to other HPC applications too.
In evaluation, we break down the execution time and compare the performance
overhead when using FTI with Restart, when using FTI with ULFM,
and when using FTI with Reinit, respectively. 
All the above experiments are running in four different scaling sizes (64 processes, 128 processes, 256 processes, and 512 processes on 32 nodes), in three different input sizes (small, medium, and large), 
and with or without injecting process failures.

In particular, our contributions are:

\begin{enumerate}

\item We present MATCH, an MPI fault tolerance benchmark suite. 
This is the first benchmark suite designed for MPI fault tolerance.
We illustrate the process and manifest the details of implementing three different fault 
tolerance designs to HPC proxy applications.

\item 
To facilitate checkpointing, we propose three principles to automatically detect data objects for checkpointing. Those data objects are the only necessary data objects to guarantee the application execution correctness after restoring the application state. 

\item 
We comparatively and extensively investigate the performance efficiency of different fault tolerance designs. 
Our evaluation reveals that, for MPI global-restart recovery, using FTI with Reinit is the most efficient design within the three evaluated fault tolerance designs, and Reinit recovery is four times faster than ULFM recovery on average, and 16 times faster than Restart on average. 


\end{enumerate}

\section{Background} 

\subsection{MATCH}
There is no existing benchmark suite aiming at benchmarking of MPI fault tolerance. We 
design, implement and test the benchmark suite MATCH to understand and comparatively study
the performance efficiency of different MPI fault tolerance designs.
MATCH is composed of six HPC proxy applications, taken from widely used HPC benchmark suites, 
aiming to 
represent the HPC application domain.
Our fault tolerance design has two interfaces: the checkpointing interface to preserve and protect 
application data, and the failure recovery interface to protect and repair the MPI communicator. 
We use the Fault Tolerance Interface (FTI) for checkpointing, and Restart, ULFM, and Reinit for MPI process 
recovery. 

\subsection{Workloads}
\label{CPR}
Our workloads comprise of proxy applications present in well-known benchmark suites: ECP proxy applications suite~\cite{richards2020quantitative} and LLNL ASC proxy applications suite~\cite{neely2017application}. Proxy applications are small and simplified applications that allow HPC practitioners,
operators, and domain scientists to explore and test key features of real applications in a quick turnaround fashion.
Our workloads represent the most important HPC application domains in scientific computing, such as iterative solvers, multi-grid, molecular dynamics, etc.
We describe the six proxy applications used in MATCH below. 

\texttt{AMG}: 
An algebraic multi-grid solver dealing with linear systems in unstructured grids problems. AMG is built on top of the BoomerAMG solver of the HYPRE library which is a large-scale linear solver library developed at LLNL. AMG provides a number of tests for a variety of problems. The default one is an anisotropy problem in the Laplace domain.

\texttt{CoMD}: 
A proxy application in Molecular Dynamics (MD) commonly used as a research platform for particle motion simulation. Different from previous MD proxy applications such as miniMD, the design of CoMD is significantly modularized which allows performing analyses on individual modules.

\texttt{LULESH}: 
A proxy application that solves the hydrodynamics equation in a Sedov blast problem.
LULESH solves the hydrodynamics equation separately by using a mesh to simulate the Sedov blast problem which is divided into a composition of volumetric elements. 
This mesh is an unstructured hex mesh, where nodes are points connected by mesh lines.

\texttt{miniFE}: 
A proxy application that solves unstructured implicit finite element problem. 
miniFE aims at the approximation of an unstructured implicit finite element.

\texttt{miniVite}: 
A proxy application that solves the graph community detection problem using the distributed Louvain method. 
The Louvain method is a greedy algorithm for the community detection problem.

\texttt{HPCCG}: 
A preconditioned conjugate gradient solver that solves the linear system of partial differential equations in a 3D chimney domain.
HPCCG approximates practical physical applications that simulate unstructured grid problems.

\subsection{Checkpointing Interface - FTI}
\label{fault tolerance frameworks}
Fault Tolerance Interface (FTI)~\cite{bautista2011fti} is an application-level, multi-level checkpointing interface for efficient
checkpointing in large-scale high-performance computing systems. 
FTI provides programmers a API, which is easy to use and the user can choose a checkpointing strategy that fits the application needs.
FTI enables multiple levels of reliability with different performance efficiency by utilizing local storage, data replication, and erasure codes. 
It requires users to designate which data objects to checkpoint, while hiding any data processing details from them. 
Users need only to pass to 
FTI the memory address and data size of the date object to be protected to enable checkpointing of this data object. 
Because failures can corrupt either a single or multiple nodes during the execution of an application,
FTI provides multiple levels of resiliency to recover from failures of different severity. Namely, the levels are the following:
\begin{itemize}
    \item 
    L1: This level stores checkpoints locally to each compute node. In a node failure, the application states cannot successfully be restored.
    \item
    L2: This level is built on top of L1 checkpointing. In this level, each application stores its checkpoint locally as well as to a neighboring node.
    
    \item 
    L3: In this level, the checkpoints are encoded by the Read-Solomon (RS) erasure code. 
    This implementation can survive the breakdown of half of the nodes within a checkpoint encoding group.
    The lost data can be restored from the RS-encoded files.
    \item 
    L4: This level flushes checkpoints to the parallel file system. This level enables differential checkpointing. 
\end{itemize}

FTI has proposed a multi-level checkpointing model, and have conducted an extensive study of correctness and reliability of this proposed checkpointing model. 
In our work, for the first time, we use FTI in the context of MPI recovery.

\subsection{Failure Recovery Interface - ULFM and Reinit}
MPI failure recovery has multiple modes, including \textit{global}, 
\textit{local}, \textit{backward}, \textit{forward}, \textit{shrinking}, 
and \textit{non-shrinking}. 

\textbf{Global}: The application execution must roll back all processes (including survivor and failed processes) to a global state to fix a failure. 

\textbf{Local}: The application can continue the execution by repairing only the failed components, such as the failed processes, to continue the execution. 

\textbf{Backward}: The application execution must go back to some previous correct state to survive a failure. 

\textbf{Forward}: The failure can be fixed with the current application state, and the execution can continue. 

\textbf{Non-shrinking}: The application needs to bring back all failed processes to resume execution.

\textbf{Shrinking}: The application execution is able to continue with the remaining survivor processes.

We target global, backward, non-shrinking recovery in this work, because this recovery fits best for the widely used Bulk Synchronous Parallel (BSP) paradigm of HPC applications. 


\subsubsection{ULFM}
User-level Fault Mitigation~\cite{bland2013post} is a leading MPI failure recovery framework providing shrinking recovery and non-shrinking recovery. 
ULFM develops new MPI operations to add fault tolerance functionalities. These functionalities include fault detection, communicator repairing, and failure recovery. 
Particularly, ULFM leverages the MPI error handler to provide notification of process failures. 
Once a failure is detected, the notified applications invokes ULFM to issue the operation MPI\_Comm\_revoke(), which revokes 
processes in the communicator. This operation interrupts any pending communication 
for this communicator for all member processes. ULFM then removes the failed processes 
using an operation MPI\_Comm\_shrink(), which creates a new communicator consisting only of survivor processes. 
Shrinking recovery is done using the steps described above.
For non-shrinking recovery, ULFM further uses the MPI\_Comm\_spawn() operation to spawn new 
processes and create a new communicator. ULFM then uses the MPI\_Intercomm\_merge() operator to merge the communicator of survivor processes and the communicator of spawned processes to create a new, combined communicator. 
We provide a sample implementation of ULFM non-shrinking recovery in Figure~\ref{fig:ulfm_impl}. 

\subsubsection{Reinit}

Reinit~\cite{doi:10.1002/cpe.4863,isc2020reinit,doi:10.1177/1094342015623623} is an alternative recovery framework designed particularly for global backward non-shrinking recovery.
Reinit implements the recovery process into the MPI runtime, thus it is transparent to users.
Therefore, the programming effort of using Reinit is much less than using ULFM. 
Programmers only need to set a global restart point. The remaining recovery is done by Reinit. 
Reinit is much more efficient than ULFM because of MPI recovery transparently handled in MPI runtime~\cite{isc2020reinit}, whereas ULFM 
recovery is handled not only in MPI runtime but also in the application. 

\section{Design}
We present design details in this section.
In particular, we describe the algorithm that we use to find data objects for 
checkpointing through data dependency analysis. 

\subsection{Find Data Objects for Checkpointing}
Unlike many fault tolerance frameworks that request programmers to decide data objects for checkpointing, 
we develop a practical analytic tool to guide programmers to identify data objects that must be checkpointed to 
recover the application execution to the same state as before the failure. 
We identify data objects for checkpointing through data dependency analysis across iterations following three \textbf{principles}.

\begin{enumerate}
    \item 
The data objects for checkpointing across iterations must be defined before 
the iterative computation. Data objects defined locally within the main computation loop are excluded from checkpointing. 

    \item
The data objects for checkpointing must be used (read or written) across iterations of the main computation loop. 
    \item
The value of data objects for checkpointing must vary across iterations of the main computation loop. 

\end{enumerate}

Following the three principles, we design and develop a data dependency analysis tool. 
The \textbf{input} to the tool is a dynamic execution instruction trace generated using LLVM-Tracer~\cite{ispass-13:shao}.
The trace contains detailed information of dynamic operations, such as the register name and memory address, the operator, 
and the line number in the source code where the operation performs. 
We describe the algorithm of the data dependency analysis tool in Algorithm~\ref{algo:checkpointing}.
The input to the algorithm is the set of locations used within the main computation loop 
and the set of locations allocated before the main computation loop. 
Those locations are either registers or memory locations. 
We create the two sets of locations by traversing the instruction trace once. 
After that, we first check values of locations and make sure the invocation values of the same location within the main computation loop 
are different. 
We then remove repetitions from both sets of locations. 
Lastly, for each location in the set of the main computation loop, we 
search for a match in the location set before the main computation loop. 
If a match is found, the matched location is used to localize data objects for checkpointing. 
The \textbf{output} of the tool is a set of locations for checkpointing.
Note that the tool only outputs the locations for checkpointing, runs separately, and does not support automatic generation of checkpointing code at this stage, which we leave as future work.


\begin{algorithm}[t]
  \caption{Find Data Objects for Checkpointing}
  \label{algo:checkpointing}
  \KwIn{$Locs\_in\_loop$: the set of locations used in the main computation loop;
   $Locs\_before\_loop$: the set of locations defined or allocated before the main computation loop}
  \KwOut{ $CPK\_Locs \colon $ the set of locations for checkpointing }
  \tcp{Check values of locations in $Locs\_in\_loop$}
  \For{$l \in Locs\_in\_loop$}{
    \eIf{The invocation values of $l$ are not the same}{
    \texttt{Keep $l$ in $Locs\_in\_loop$}\;
    }{
    \texttt{Remove $l$ from $Locs\_in\_loop$}\;
    }
  }  
  \tcp{Remove repetition in $Locs\_in\_loop$ and $Locs\_before\_loop$}
  \For{$l \in Locs\_in\_loop$}{
    Remove repetition\;
  }
  \For{$l \in Locs\_before\_loop$}{
    Remove repetition\;
  }
  \tcp{Check if locations in $Locs\_in\_loop$ can find a match in $Locs\_before\_loop$}
  \For{ $l_{i} \in Locs\_in\_loop$ } {
    \For{ $l_{j} \in Locs\_before\_loop$ } {
        \If{ $l_{i}$ matches $l_{j}$ }{
            $CPK\_Locs \leftarrow l_{i}$\;
        }
    }
  }
\end{algorithm}

\section{Implementation}
\label{sec:impl}
\subsection{FTI Implementation}
The Fault Tolerance Interface (FTI) is a checkpointing library 
widely used by HPC developers for checkpointing.
We illustrate a sample usage of FTI in Figure~\ref{fig:fti_impl}.
\emph{Please read the FTI paper~\cite{6114441} for the implementation details of FTI function calls such as FTI\_Protect() and FTI\_Recover().}
We reckon a challenge while implementing checkpointing using FTI for MATCH workloads. 

The challenge is the programming complexity of enabling FTI checkpointing to data objects when the number of data objects for checkpointing is large.
FTI requests users to add FTI checkpointing to every data object manually. 
This significantly increases the programming effort when the number of data objects for checkpointing
is large and when the data object is a complicated data structure. 
This is a common issue in application-level checkpoint libraries such as FTI, VeloC, and SCR. 
These libraries cannot automatically enable checkpointing to target data objects.

\begin{figure}
\begin{lstlisting}[xleftmargin=.02\textwidth, 
xrightmargin=.02\textwidth]
int main(int argc, char *argv[]) {
    MPI_Init(&argc, &argv);
    
    // Initialize FTI
    FTI_Init(argv[1], MPI_COMM_WORLD);

    // Right before the main computation loop
    // Add FTI protection to data objects
    FTI_Protect();
    
    // the main computation loop
    while (...) {
        // At the beginning of the loop
        // If the execution is a restart
        if ( FTI_Status() != 0){
            FTI_Recover();
        }
    
        // do FTI checkpointing
        if (Iter_Num % cp_stride == 0) {
            FTI_Checkpoint();
        }
    }
    
    FTI_Finalize();
    MPI_Finalize();
}
\end{lstlisting}
\caption{A sample implementation of FTI.}
\label{fig:fti_impl}
\end{figure}

\subsection{FTI with Reinit Implementation}
Reinit is the state-of-the-art MPI global non-shrinking recovery framework. 
Reinit hides all of the recovery implementation in the MPI runtime, which makes it ease-to-use. 
We provide a sample implementation of Reinit with FTI checkpointing in Figure~\ref{fig:reinit_impl}.
We can see that Reinit recovery only adds less than five lines of code. 
Line 4 and 5 are for Reinit recovery, while Line 14 is used for other functions. 
FTI is completely independent of Reinit. To implement FTI with Reinit, 
the only thing to notice is to move the FTI\_Init() and FTI\_Finalize() 
functions into the resilient\_main() function as well.  
\emph{Please read work on Reinit~\cite{doi:10.1002/cpe.4863,isc2020reinit,doi:10.1177/1094342015623623} for the design and implementation details of Reinit.}

\begin{figure}
\begin{lstlisting}[xleftmargin=.02\textwidth, 
xrightmargin=.02\textwidth]
int main(int argc, char *argv[])
{
    MPI_Init(&argc, &argv);
    OMPI_Reinit(argc, argv, resilient_main);
    MPI_Finalize();
    return 0;
}
// Move the original main() into resilient_main()
int resilient_main(int argc, char** argv, OMPI_reinit_state_t state) {
    FTI_Init(argv[1], MPI_COMM_WORLD);
    ...
    // the main computation loop
    ...
    FTI_Finalize();
    return 0;
}
\end{lstlisting}
\caption{A sample implementation of Reinit.}
\label{fig:reinit_impl}
\end{figure}

\subsection{FTI with ULFM Implementation}
ULFM is a pioneer MPI recovery framework. ULFM provides five 
new MPI interfaces to support MPI fault tolerance. ULFM gives the flexibility to programmers to use provided 
interfaces to implement their own, customized MPI recovery strategy. Also, ULFM allows 
programmers to use both shrinking and non-shrinking recovery. However, it takes a significant learning and 
programming effort before a programmer can successfully implement recovery with ULFM.
As most HPC applications follow the Bulk Synchronous Parallel (BSP) paradigm, 
we focus on ULFM global non-shrinking recovery. In order to implement ULFM non-shrinking recovery, we add 
more than 200 lines of code for each benchmark, which requires more effort compared to 
the implementation (less than five lines of code) using Reinit for recovery.
We provide a sample 
implementation of ULFM global non-shrinking recovery with FTI in Figure~\ref{fig:ulfm_impl}. 

\begin{figure}[h]
\begin{lstlisting}[xleftmargin=.02\textwidth, xrightmargin=.02\textwidth]
/* world will swap between worldc[0] and worldc[1] after each respawn */
MPI_Comm worldc[2] = { MPI_COMM_NULL, MPI_COMM_NULL };
int worldi = 0;

//the MPI communicator must be implemented as a global variable to enable immediately update after ULFM recovery for FTI to use
#define world (worldc[worldi]) 

int main(int argc, char *argv[])
{
    MPI_Init(&argc, &argv);
    // set long jump
    int do_recover = _setjmp(stack_jmp_buf);
    int survivor = IsSurvivor();
    /* set an errhandler on world, so that a failure is not fatal anymore */
    MPI_Comm_set_errhandler(world);
    FTI_Init(argv[1], world);
    ... 
    // the main computation loop
    ...
    FTI_Finalize();
    MPI_Finalize(); 
}

/* error handler: repair comm world */
static void errhandler(MPI_Comm* pcomm, int* errcode, ...)
{
    int eclass;
    MPI_Error_class(*errcode, &eclass);

    if( MPIX_ERR_PROC_FAILED != eclass &&
        MPIX_ERR_REVOKED != eclass ) {
        MPI_Abort(MPI_COMM_WORLD, *errcode);
    }
    
    /* swap the worlds */
    worldi = (worldi+1)%2;
    
    MPIX_Comm_revoke(world);
    MPIX_Comm_shrink();
    MPI_Comm_spawn();
    MPI_Intercomm_merge();
    MPIX_Comm_agree();
    
    _longjmp( stack_jmp_buf, 1 );
}
\end{lstlisting}
\caption{A sample implementation of ULFM non-shrinking recovery.}
\squeezeup
\label{fig:ulfm_impl}
\end{figure}

When combining ULFM global non-shrinking recovery with FTI, it is important to notice that the MPI\_COMM\_WORLD 
must be implemented as a global variable with an external declaration. 
See Lines 2-6 in Figure~\ref{fig:ulfm_impl} for the implementation details.
This is because ULFM updates the world communicator handler
and FTI must use the repaired world communicator for MPI communication to correctly function.

\subsection{Fault Injection}
We emulate MPI process failures through fault injection. In particular, 
we raise a SIGTERM signal at a \emph{randomly} selected MPI process in a \emph{randomly} selected iteration of the 
main computation loop.
We illustrate the fault injection code in Figure~\ref{fig:fi_impl}.
Note that we choose to evaluate different fault tolerance techniques by triggering a process failure, 
which does not mean that the MPI recovery frameworks do not support recovery from a node failure. 
On the one hande Reinit can recover from a node failure~\cite{isc2020reinit}, on the other hand the current ULFM 
implementation cannot.  
In our case, it is sufficient to evaluate on MPI process failures to compare the performance difference when using FTI checkpointing in ULFM and Reinit. 


\begin{figure}
\begin{lstlisting}[xleftmargin=.02\textwidth, 
xrightmargin=.02\textwidth]
// simulation of proc failures
if (procfi == 1 && numIters==Selected_Iter){
    if (myrank == Selected_Rank){
       printf("KILL rank %d\n", myrank);
       kill(getpid(), SIGTERM);
    }
}
\end{lstlisting}
\vspace{-10pt}
\caption{A sample implementation of fault injection.}
\label{fig:fi_impl}
\end{figure}
\section{Evaluation}
\label{sec:eval}

We measure and compare the MPI failure recovery time, the checkpointing time, and the application execution time of three fault tolerance designs.
We use a similar methodology with other works~\cite{isc2020reinit,bland2015lessons,adam2019checkpoint,shahzad2018craft} to evaluate MPI fault tolerance, we validate our benchmark suite MATCH on four scaling sizes and three input problem sizes, both with and without fault injection.
We answer the following questions:

\begin{itemize}
    \item Can fault tolerance interfaces (such as ULFM) delay the application execution or not? 
    
    \item Can the checkpointing interface and the MPI recovery interface interfere with each other? 
    
    \item Can ULFM perform better or Reinit perform better for different scaling sizes and different input problem sizes?
\end{itemize}

\begin{figure*}[h]
  \centering
  \begin{subfigure}[t]{.30\textwidth}
    \includegraphics[width=1.0\textwidth]{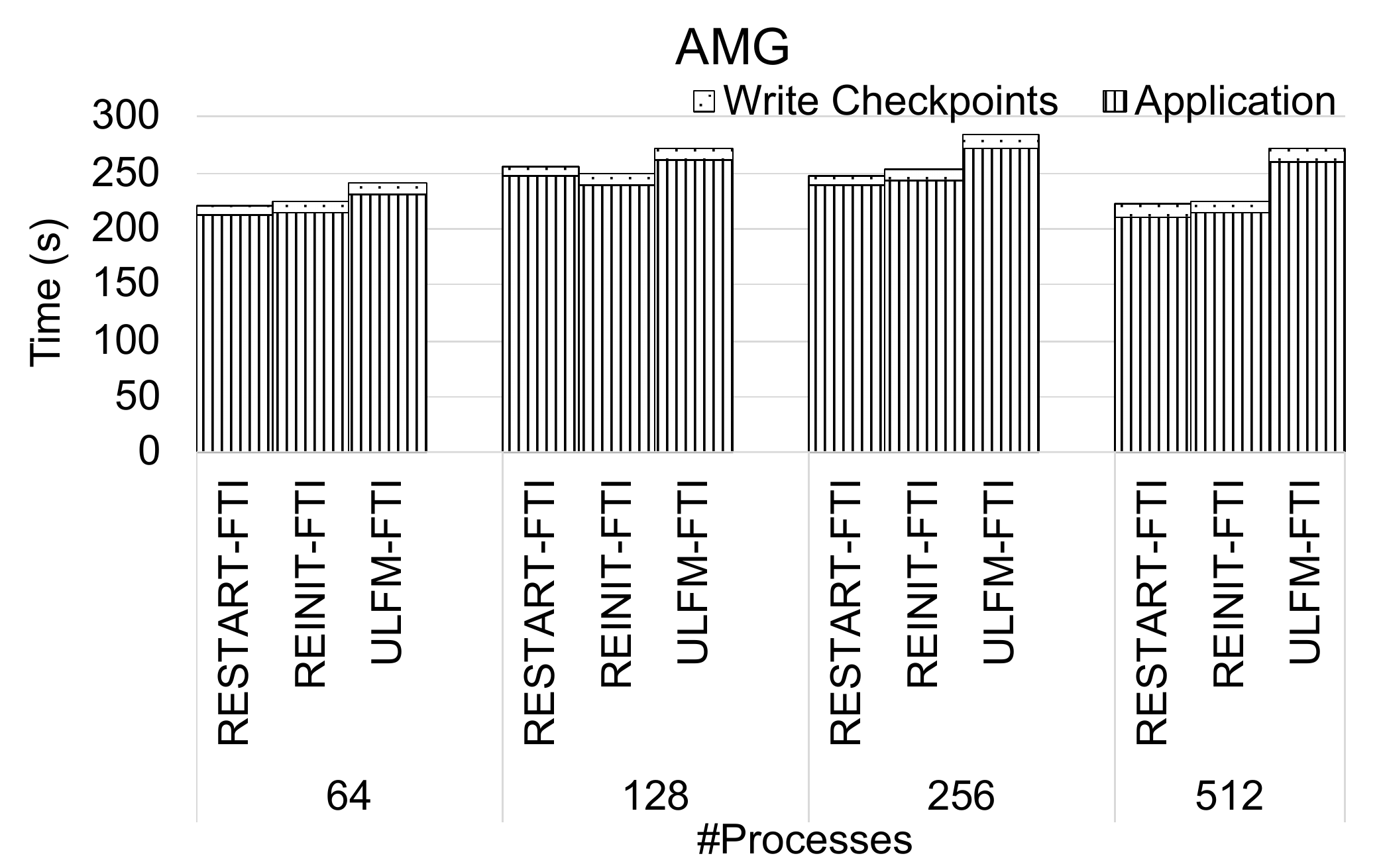}
    \caption{AMG}
  \end{subfigure}
  \begin{subfigure}[t]{.30\textwidth}
    \includegraphics[width=1.0\textwidth]{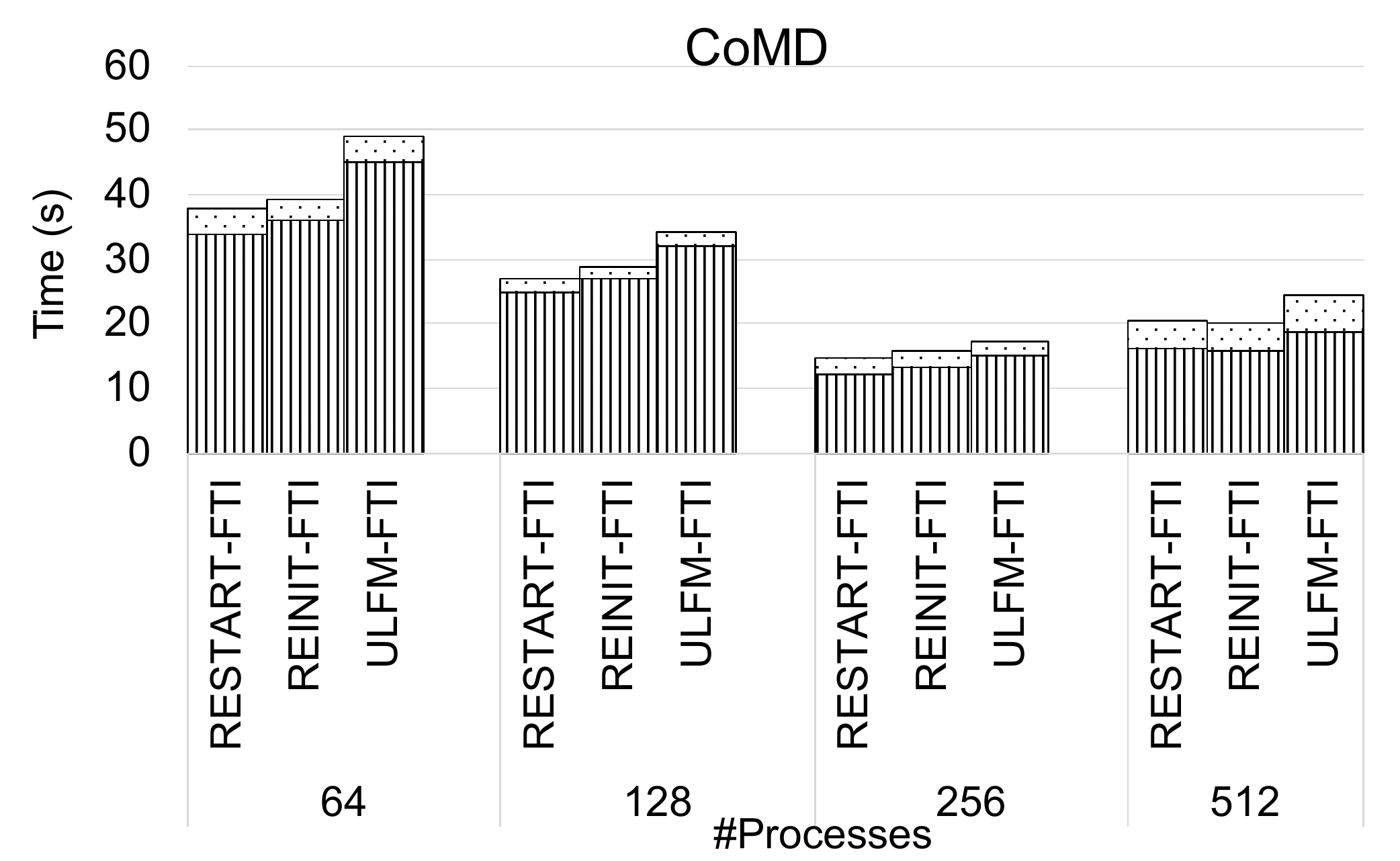}
    \caption{CoMD}
  \end{subfigure}
  \begin{subfigure}[t]{.30\textwidth}
    \includegraphics[width=1.0\textwidth]{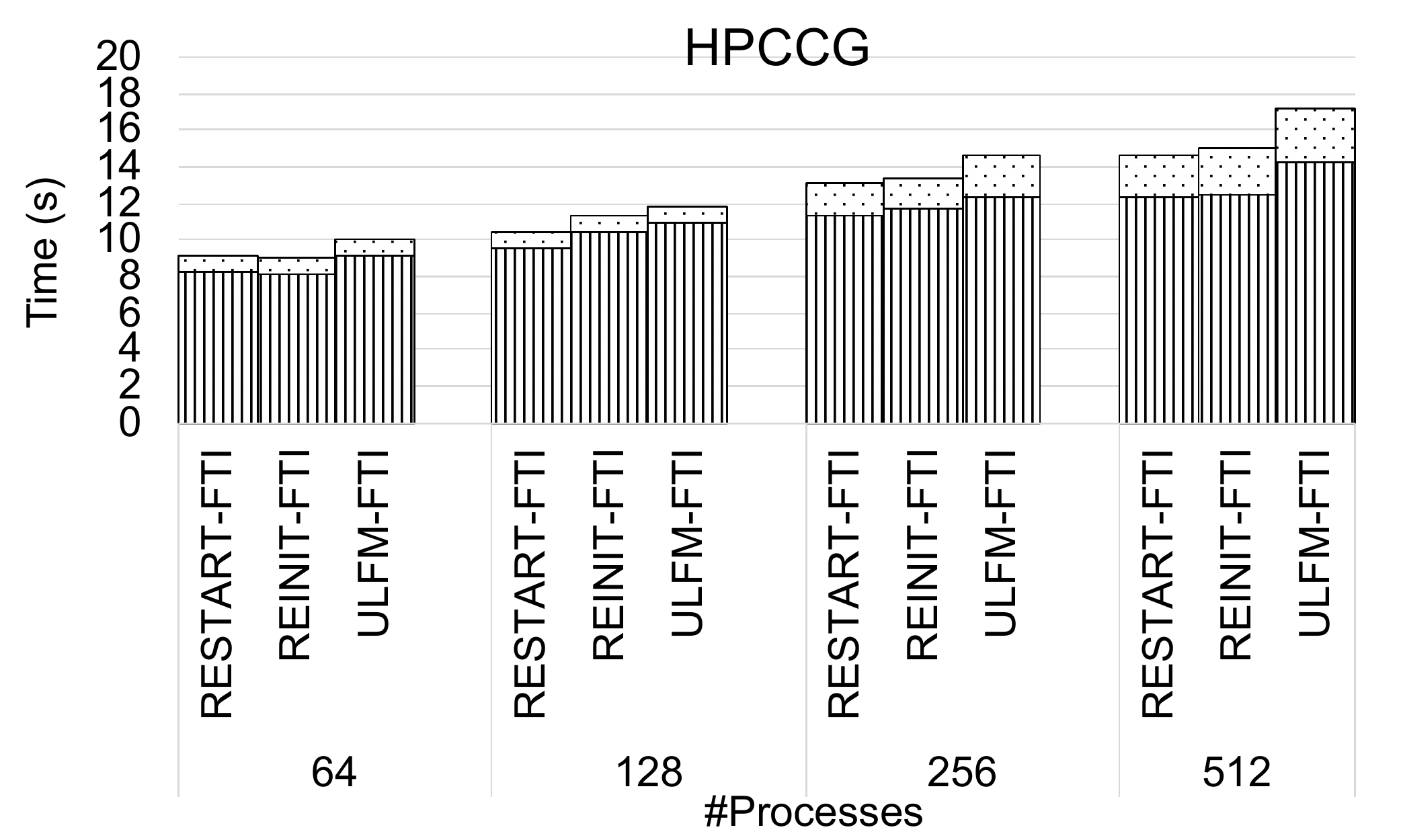}
    \caption{HPCCG}
  \end{subfigure}
    \begin{subfigure}[t]{.30\textwidth}
    \includegraphics[width=1.0\textwidth]{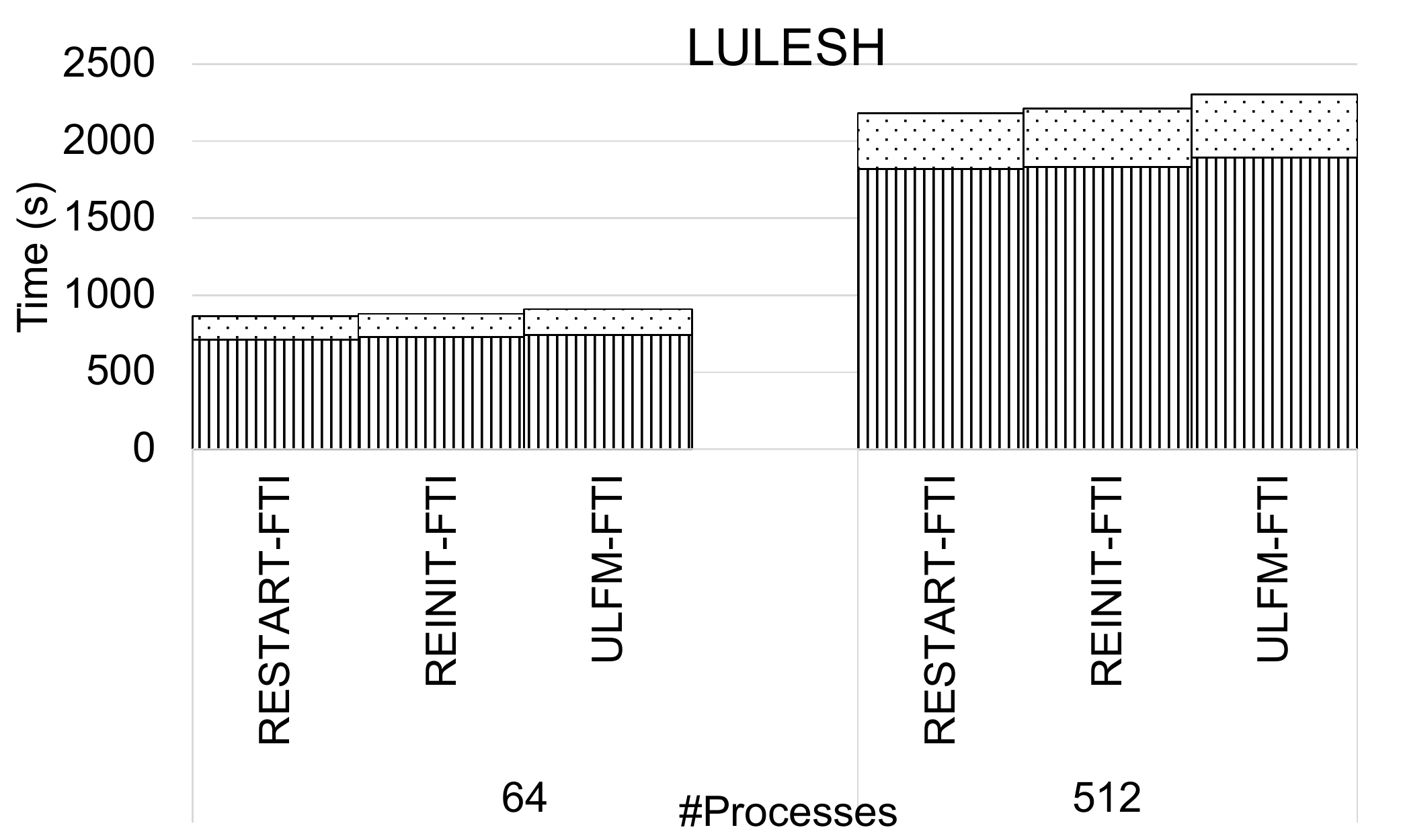}
    \caption{LULESH}
  \end{subfigure}
  \begin{subfigure}[t]{.30\textwidth}
    \includegraphics[width=1.0\textwidth]{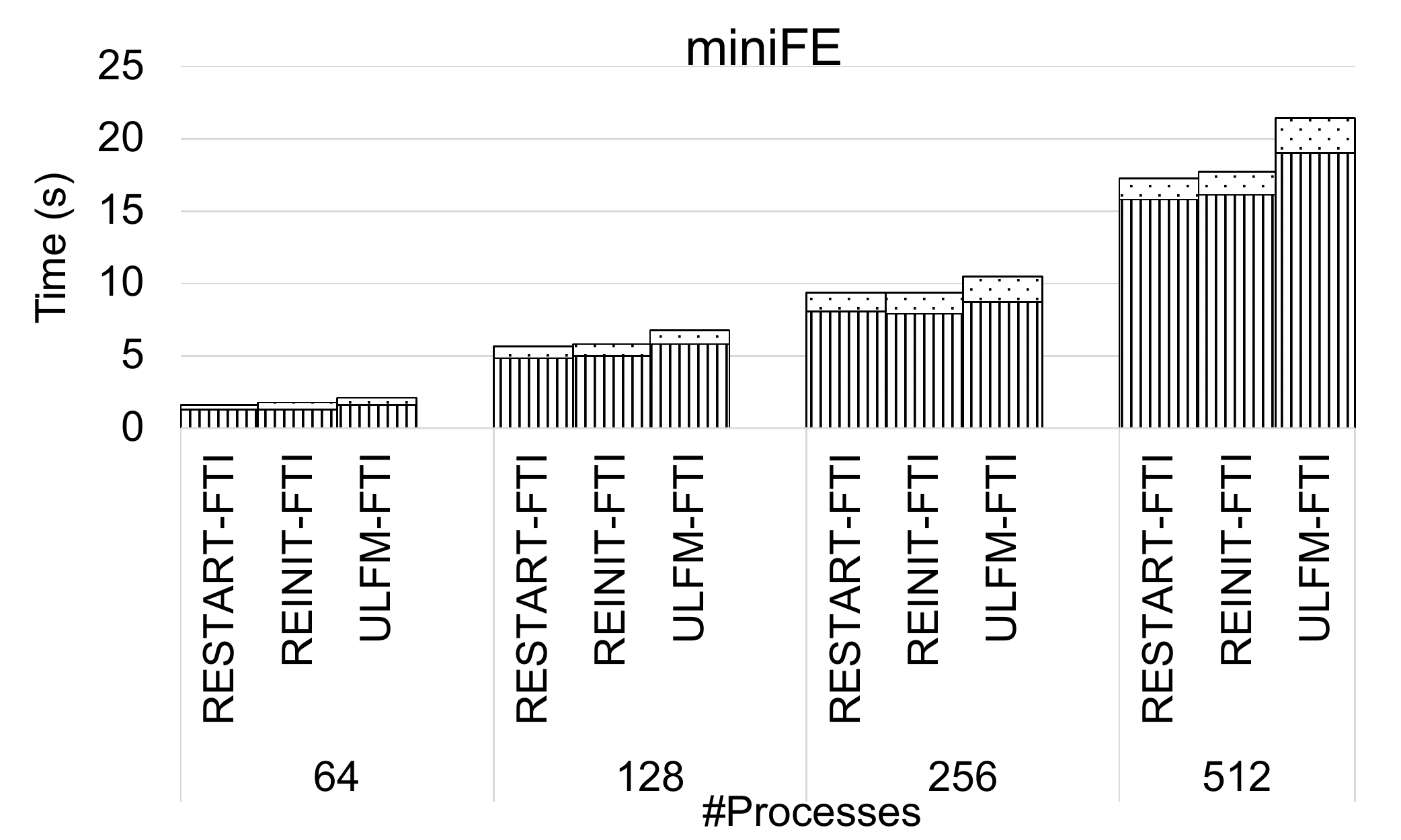}
    \caption{miniFE}
  \end{subfigure}
  \begin{subfigure}[t]{.30\textwidth}
    \includegraphics[width=1.0\textwidth]{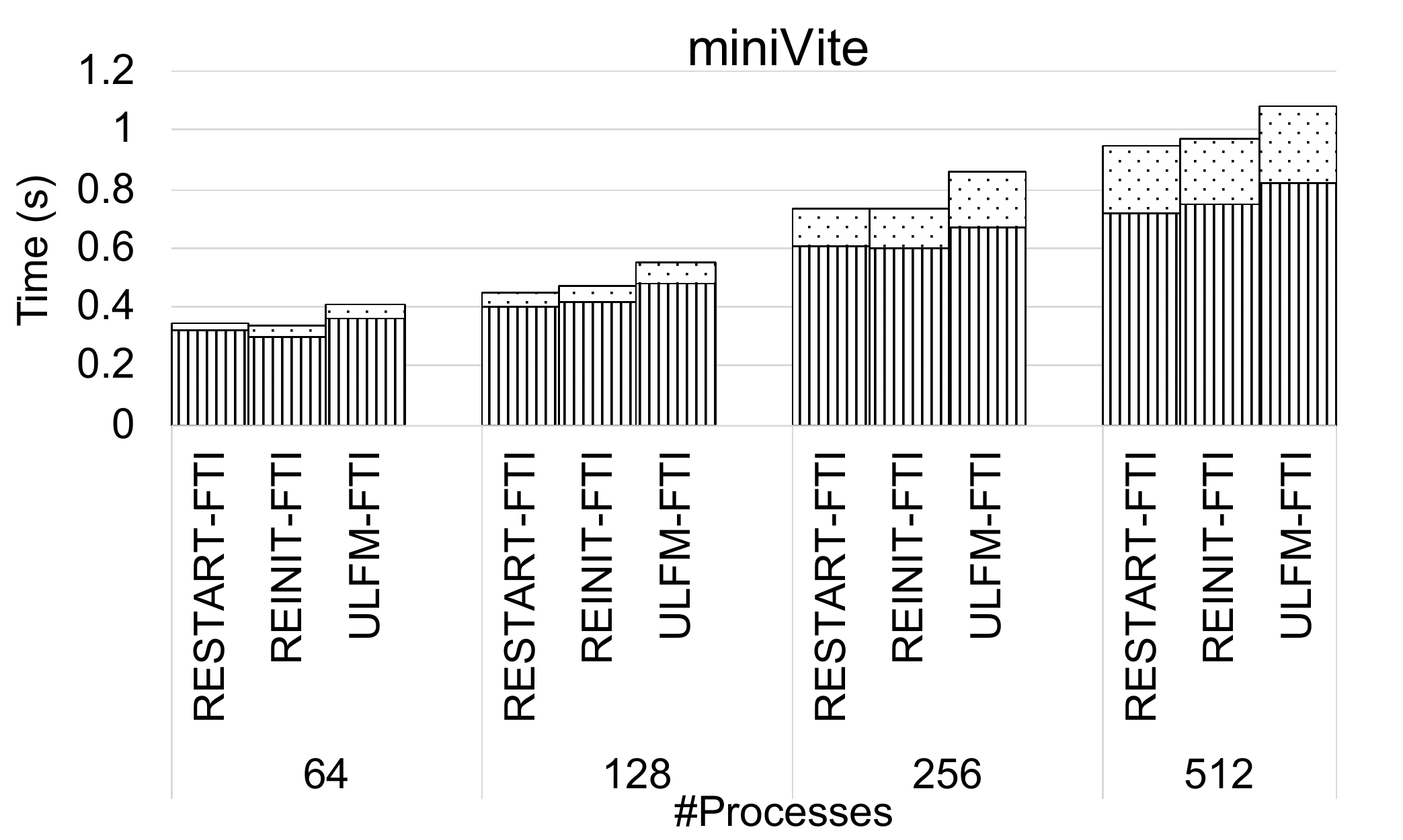}
    \caption{miniVite}
  \end{subfigure}
  \caption{Execution time breakdown recovering in different scaling sizes with no process failures}
  \label{fig:no_failure_diff_scale}
\end{figure*}

\begin{table*}[h]
  \setlength{\tabcolsep}{5pt}
  \centering
  \caption{Experimentation configuration for proxy applications (\textbf{default scaling size: 64 processes; default input problem: small})}
  \begin{tabular}{lcccc}
  \toprule
  \textbf{Application} & \textbf{Small Input} & \textbf{Medium Input} & \textbf{Large Input} & \textbf{Number of processes}\\
  \midrule
  AMG & -problem 2 -n 20 20 20 & -problem 2 -n 40 40 40 & -problem 2 -n 60 60 60 & 64, 128, 256, 512 \\
  CoMD & -nx 128 -ny 128 -nz 128 &-nx 256 -ny 256 -nz 256 &-nx 512 -ny 512 -nz 512 & 64, 128, 256, 512 \\
  HPCCG & 64 64 64 & 128 128 128 & 192 192 192 & 64, 128, 256, 512 \\
  LULESH & -s 30 -p & -s 40 -p & -s 50 -p & 64, 512 \\
  miniFE & -nx 20 -ny 20 -nz 20 & -nx 40 -ny 40 -nz 40 & -nx 60 -ny 60 -nz 60 & 64, 128, 256, 512 \\
  miniVite & -p 3 -l -n 128000 & -p 3 -l -n 256000 & -p 3 -l -n 512000 & 64, 128, 256, 512 \\
  \bottomrule
  \end{tabular}
  \label{tab:config}
\end{table*}

\begin{figure*}[h]
  \centering
  \begin{subfigure}[t]{.30\textwidth}
    \includegraphics[width=1.0\textwidth]{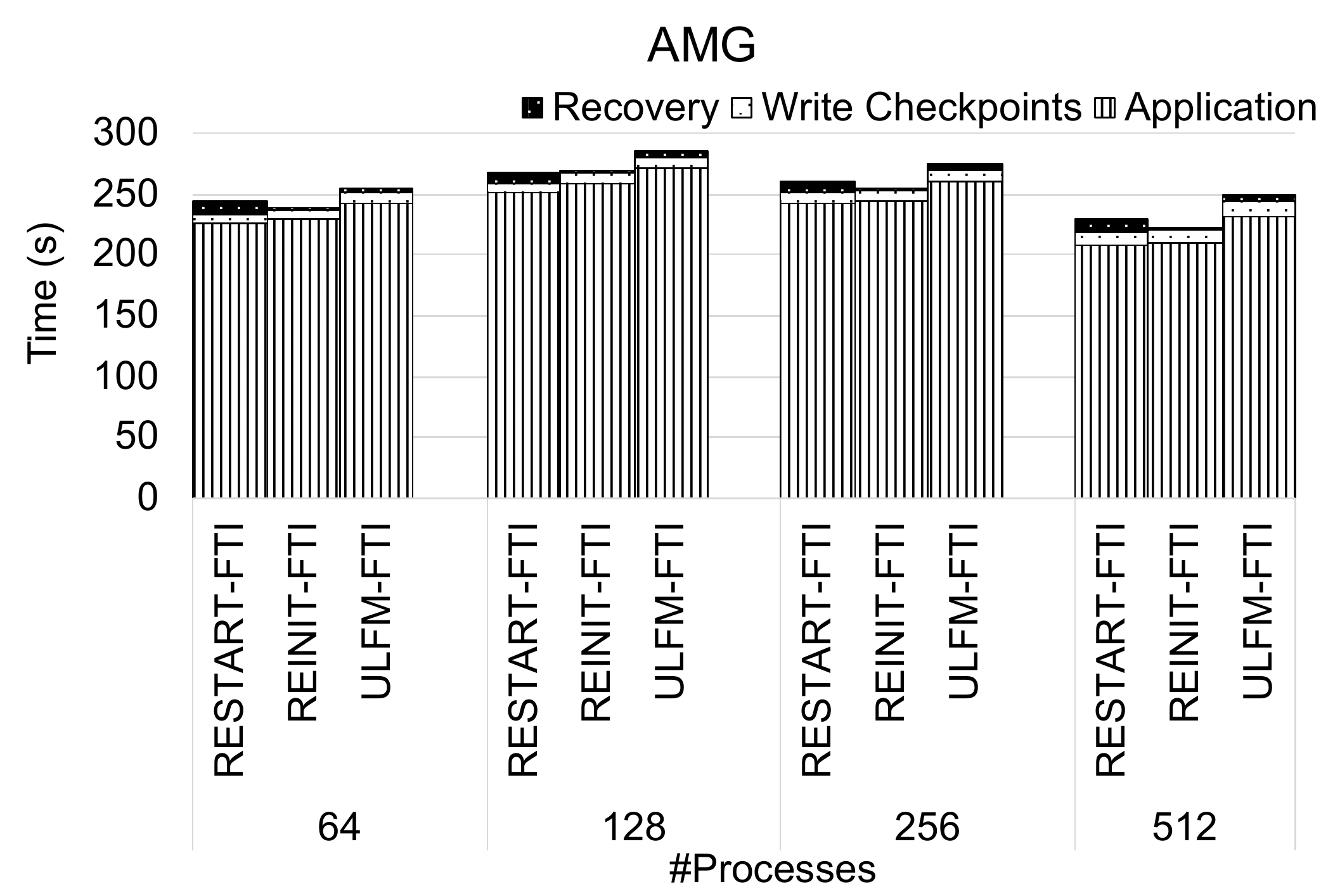}
    \caption{AMG}
  \end{subfigure}
  \begin{subfigure}[t]{.30\textwidth}
    \includegraphics[width=1.0\textwidth]{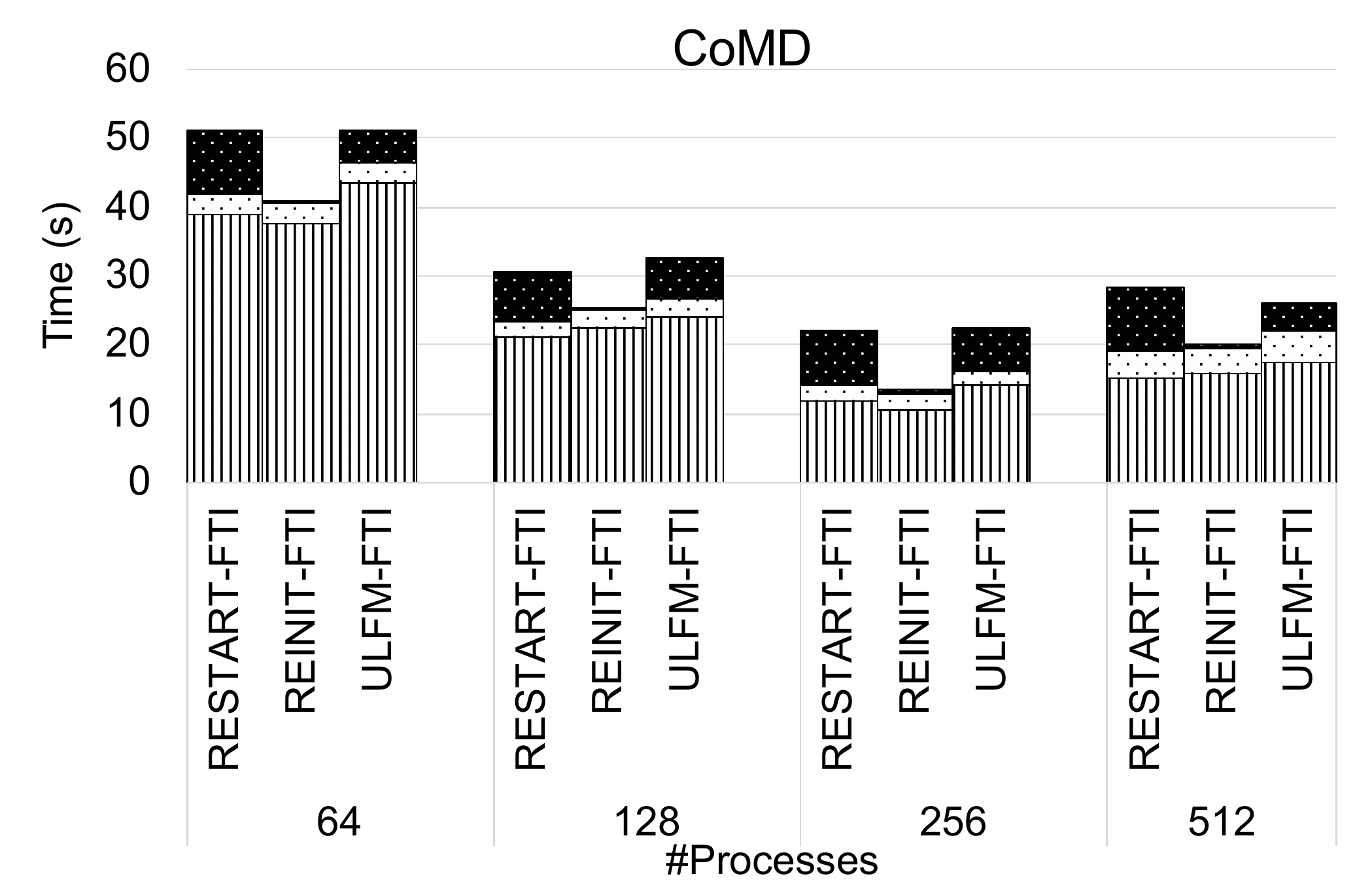}
    \caption{CoMD}
  \end{subfigure}
  \begin{subfigure}[t]{.30\textwidth}
    \includegraphics[width=1.0\textwidth]{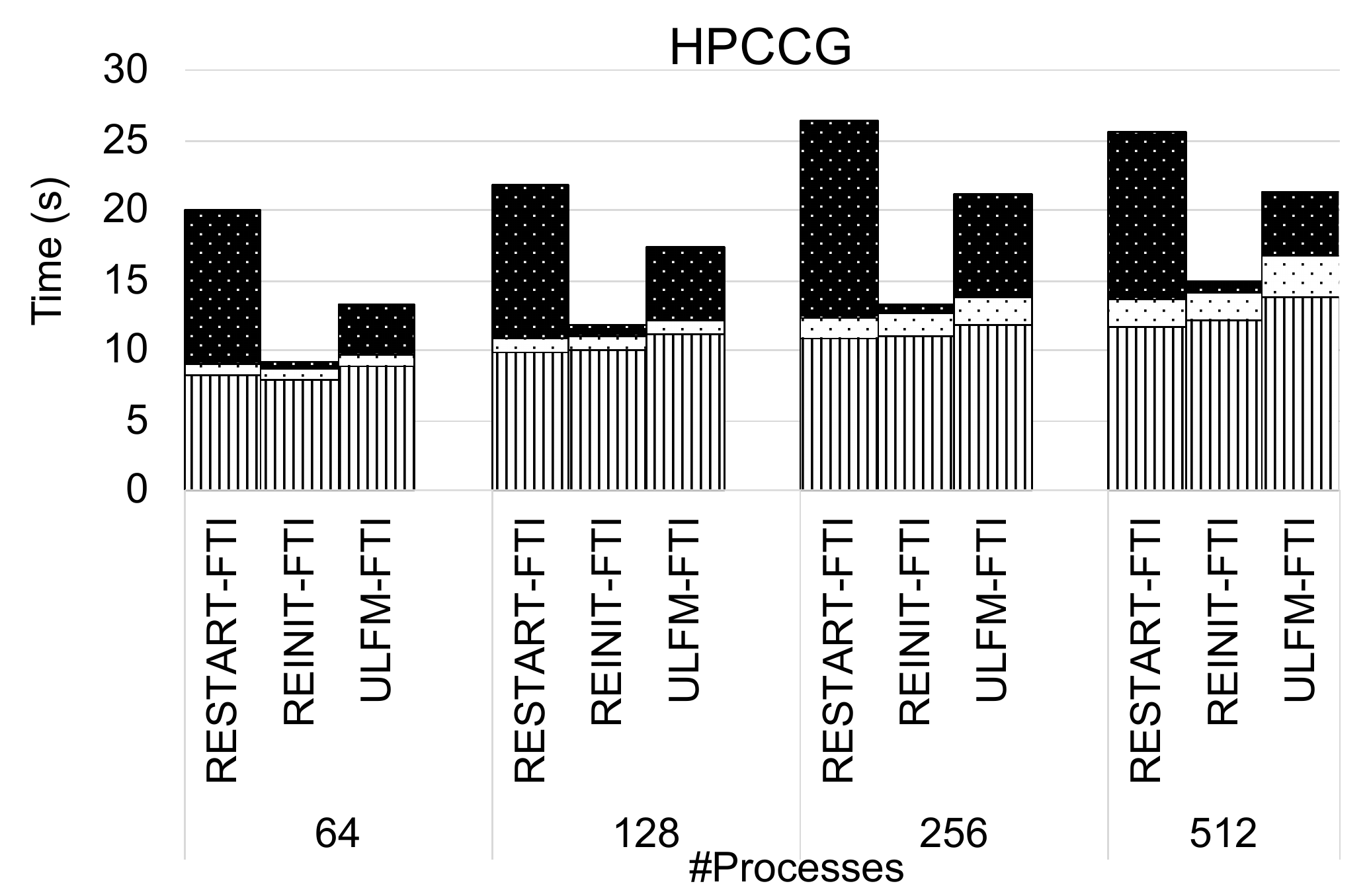}
    \caption{HPCCG}
  \end{subfigure}
    \begin{subfigure}[t]{.30\textwidth}
    \includegraphics[width=1.0\textwidth]{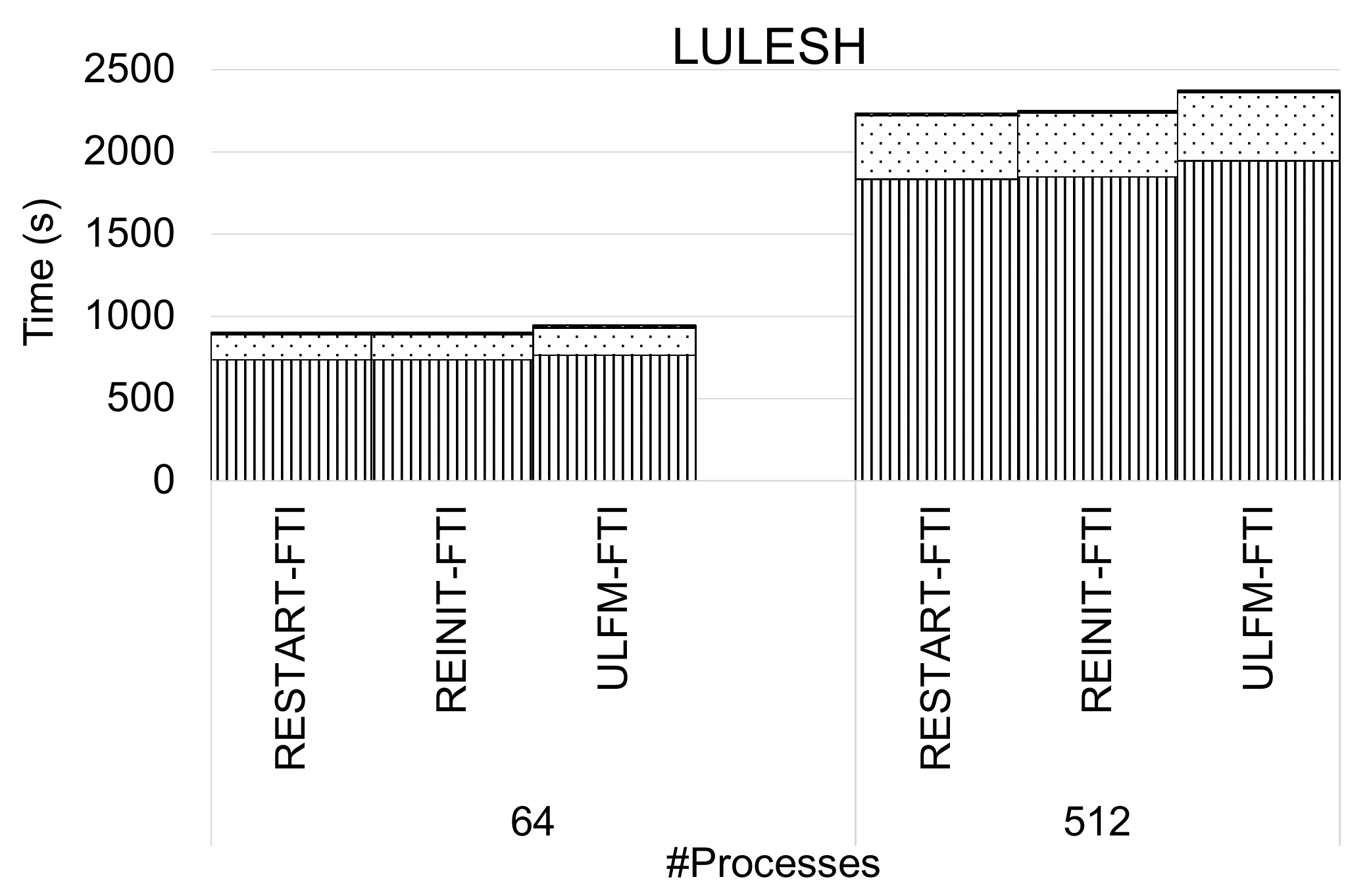}
    \caption{LULESH}
  \end{subfigure}
  \begin{subfigure}[t]{.30\textwidth}
    \includegraphics[width=1.0\textwidth]{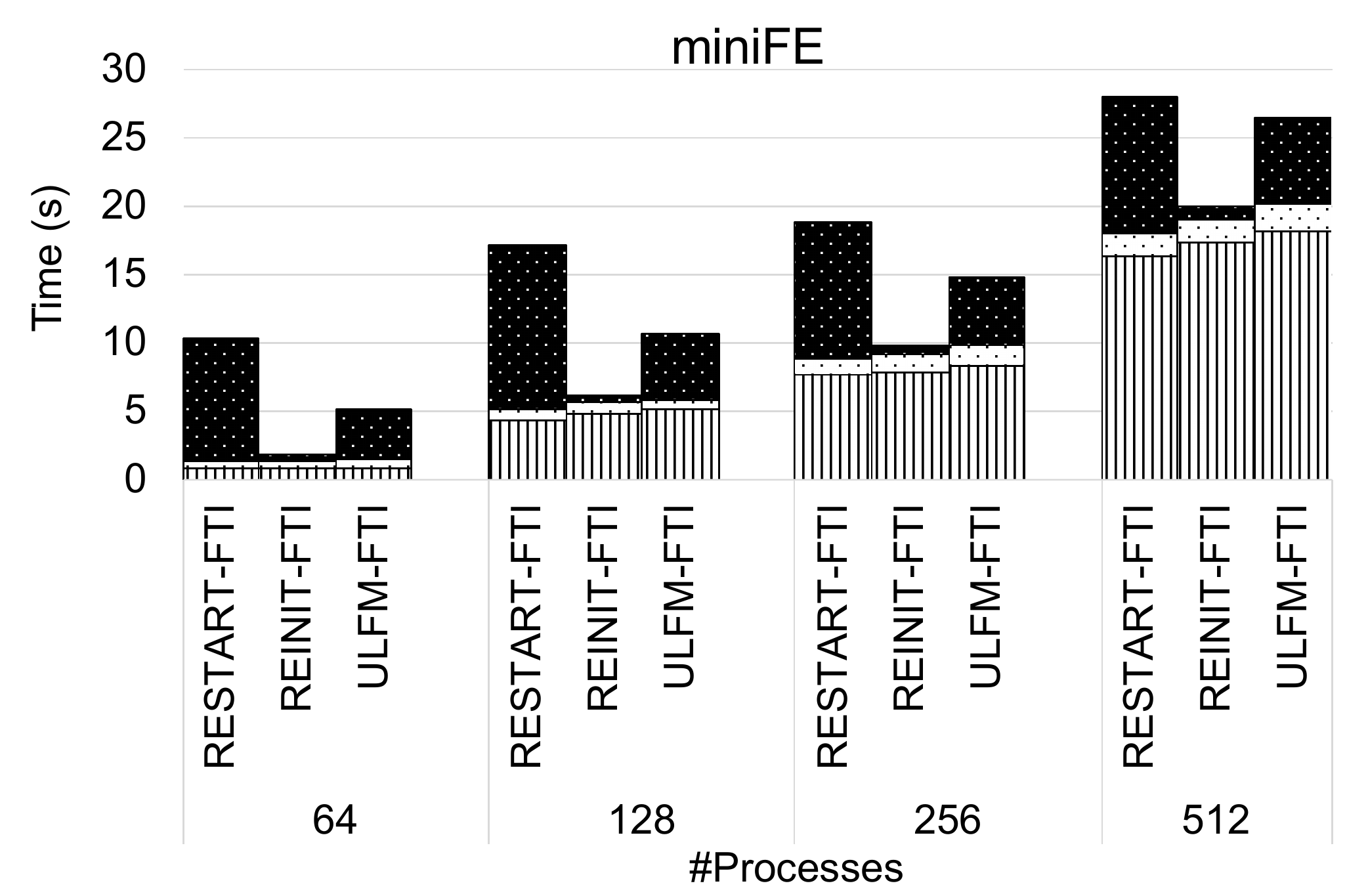}
    \caption{miniFE}
  \end{subfigure}
  \begin{subfigure}[t]{.30\textwidth}
    \includegraphics[width=1.0\textwidth]{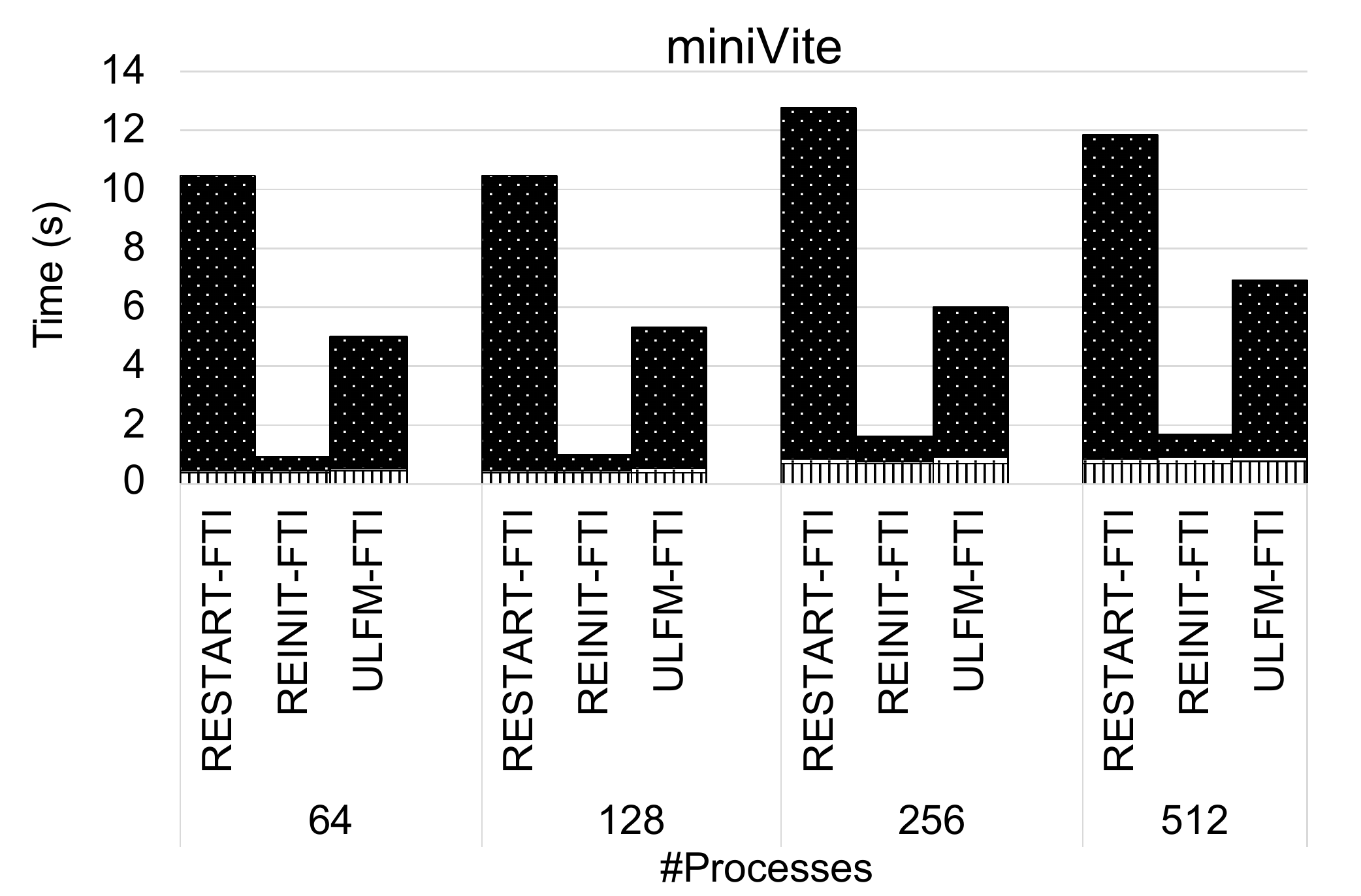}
    \caption{miniVite}
  \end{subfigure}
  \caption{Execution time breakdown recovering from a process failure in different scaling sizes}
  \label{fig:failure_diff_scale}
\end{figure*}

\begin{figure*}[h]
  \centering
  \begin{subfigure}[t]{.30\textwidth}
    \includegraphics[width=1.0\textwidth]{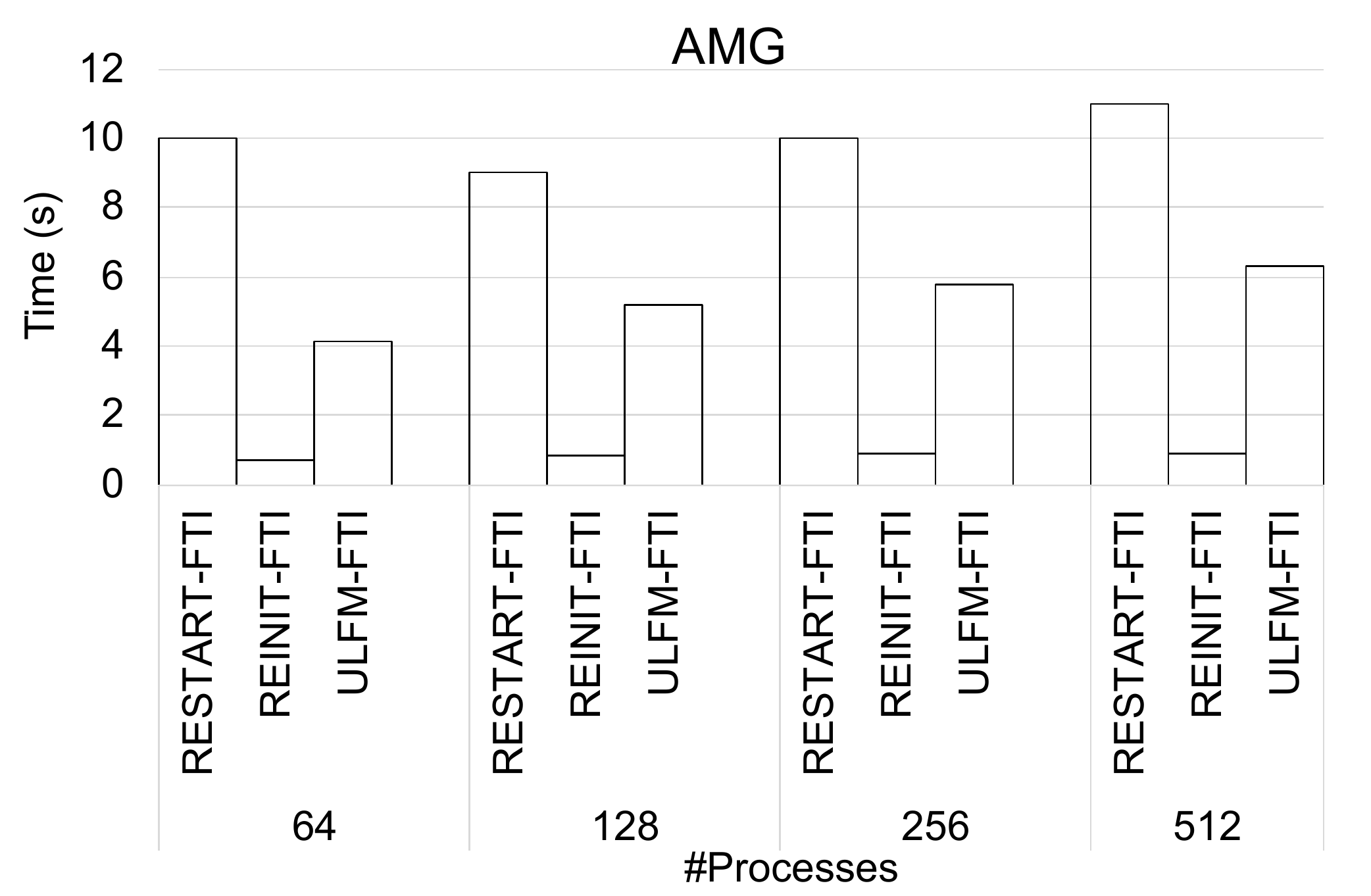}
    \caption{AMG}
  \end{subfigure}
  \begin{subfigure}[t]{.30\textwidth}
    \includegraphics[width=1.0\textwidth]{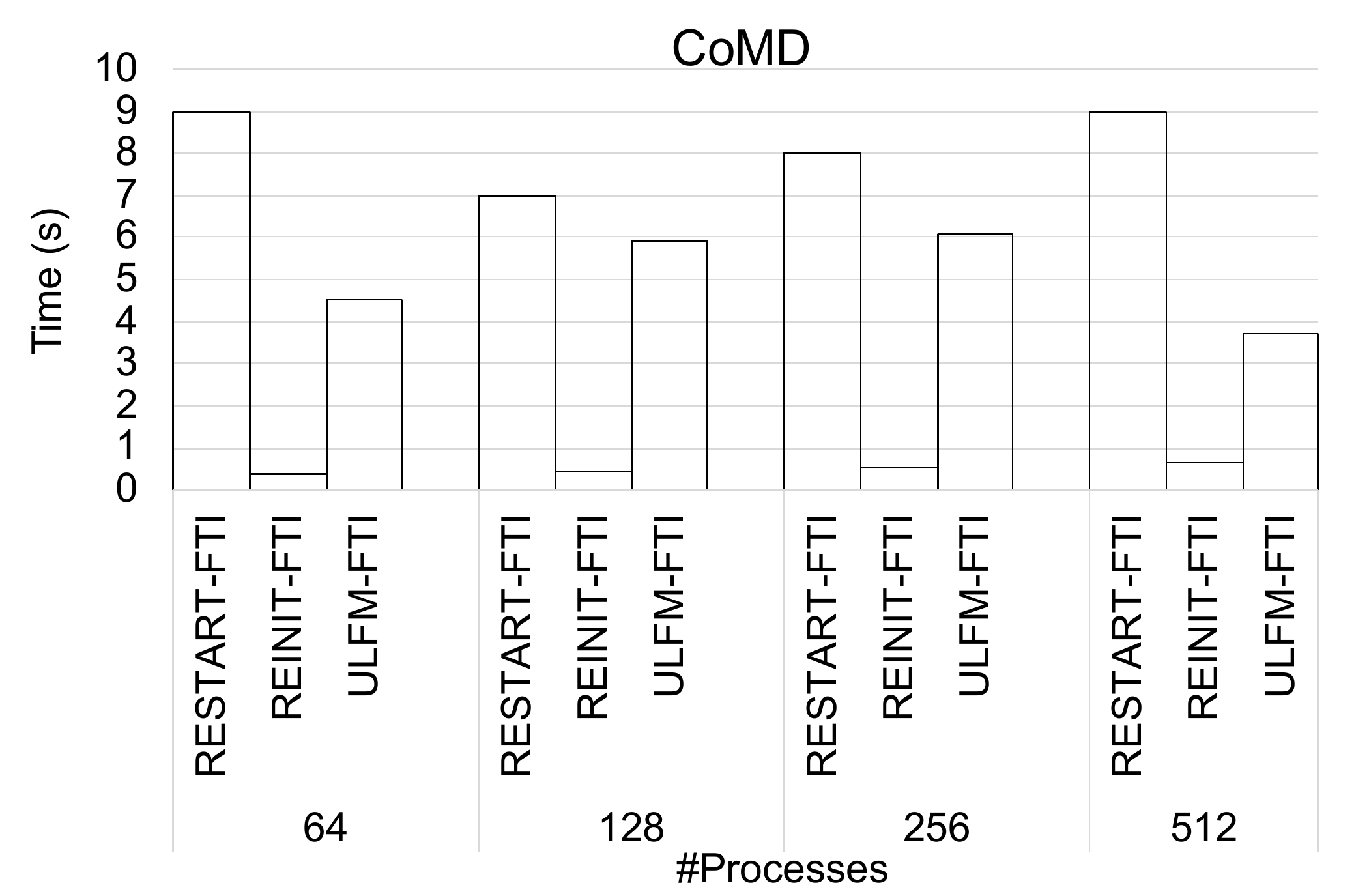}
    \caption{CoMD}
  \end{subfigure}
  \begin{subfigure}[t]{.30\textwidth}
    \includegraphics[width=1.0\textwidth]{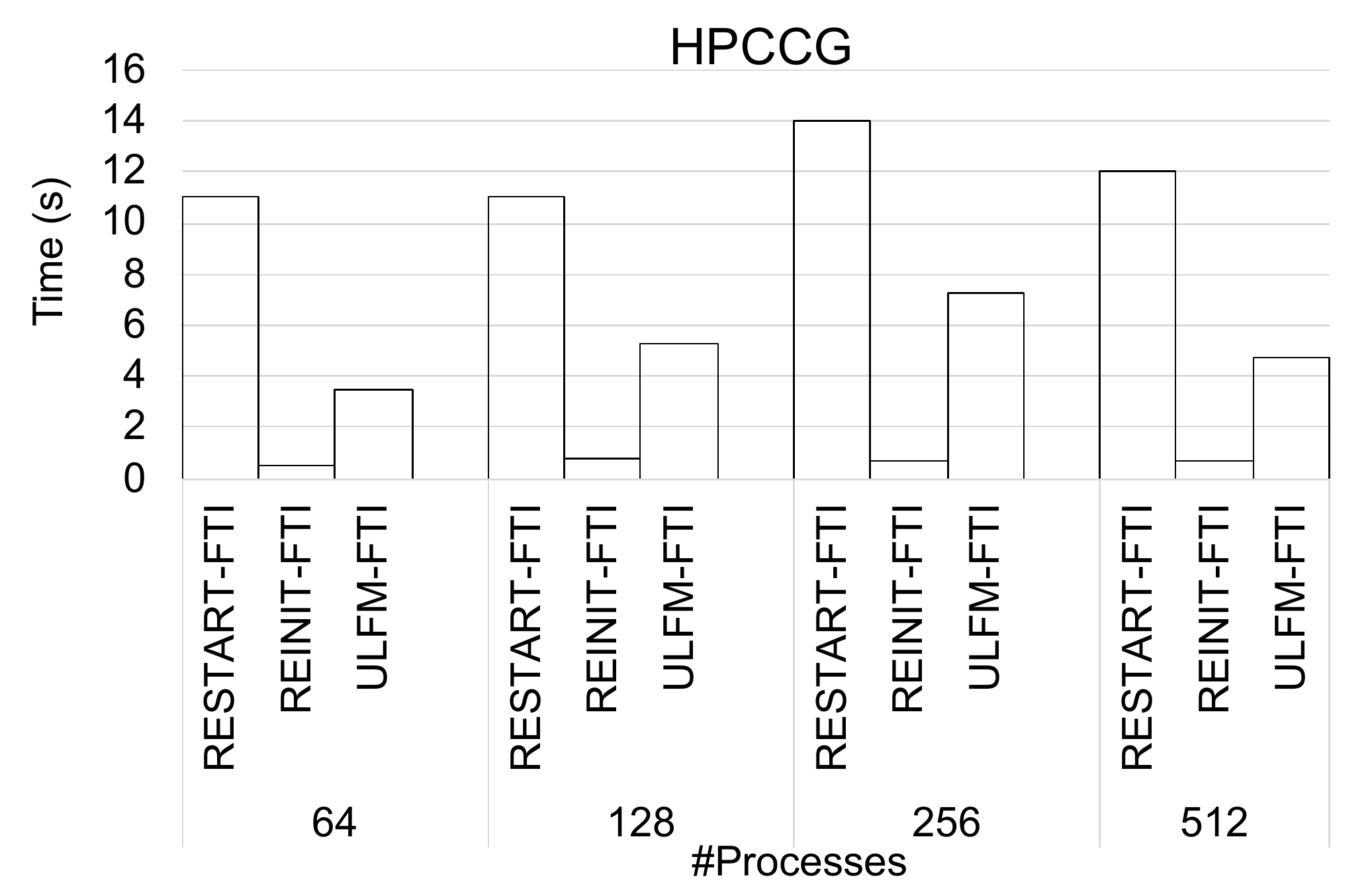}
    \caption{HPCCG}
  \end{subfigure}
    \begin{subfigure}[t]{.30\textwidth}
    \includegraphics[width=1.0\textwidth]{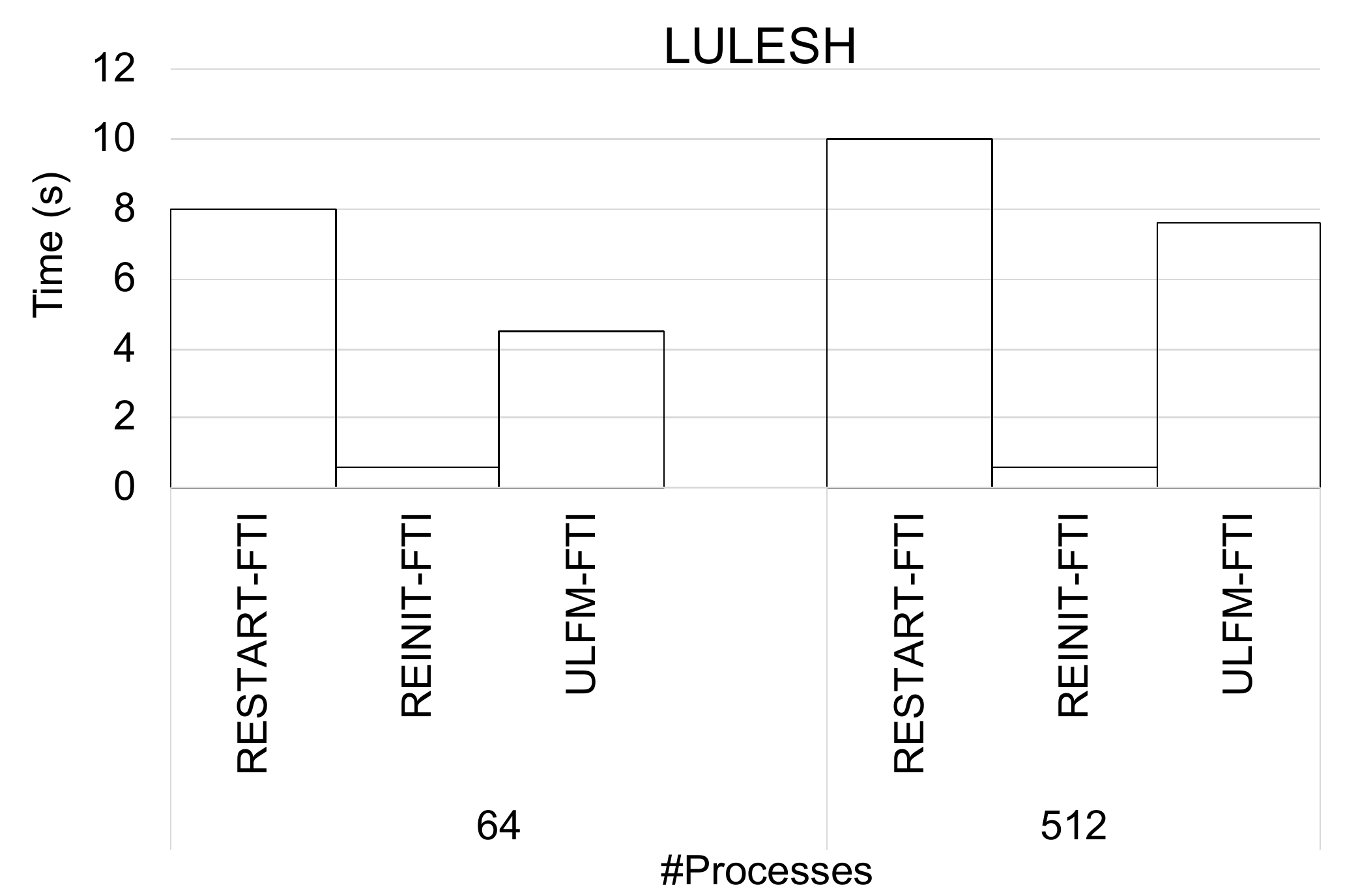}
    \caption{LULESH}
  \end{subfigure}
  \begin{subfigure}[t]{.30\textwidth}
    \includegraphics[width=1.0\textwidth]{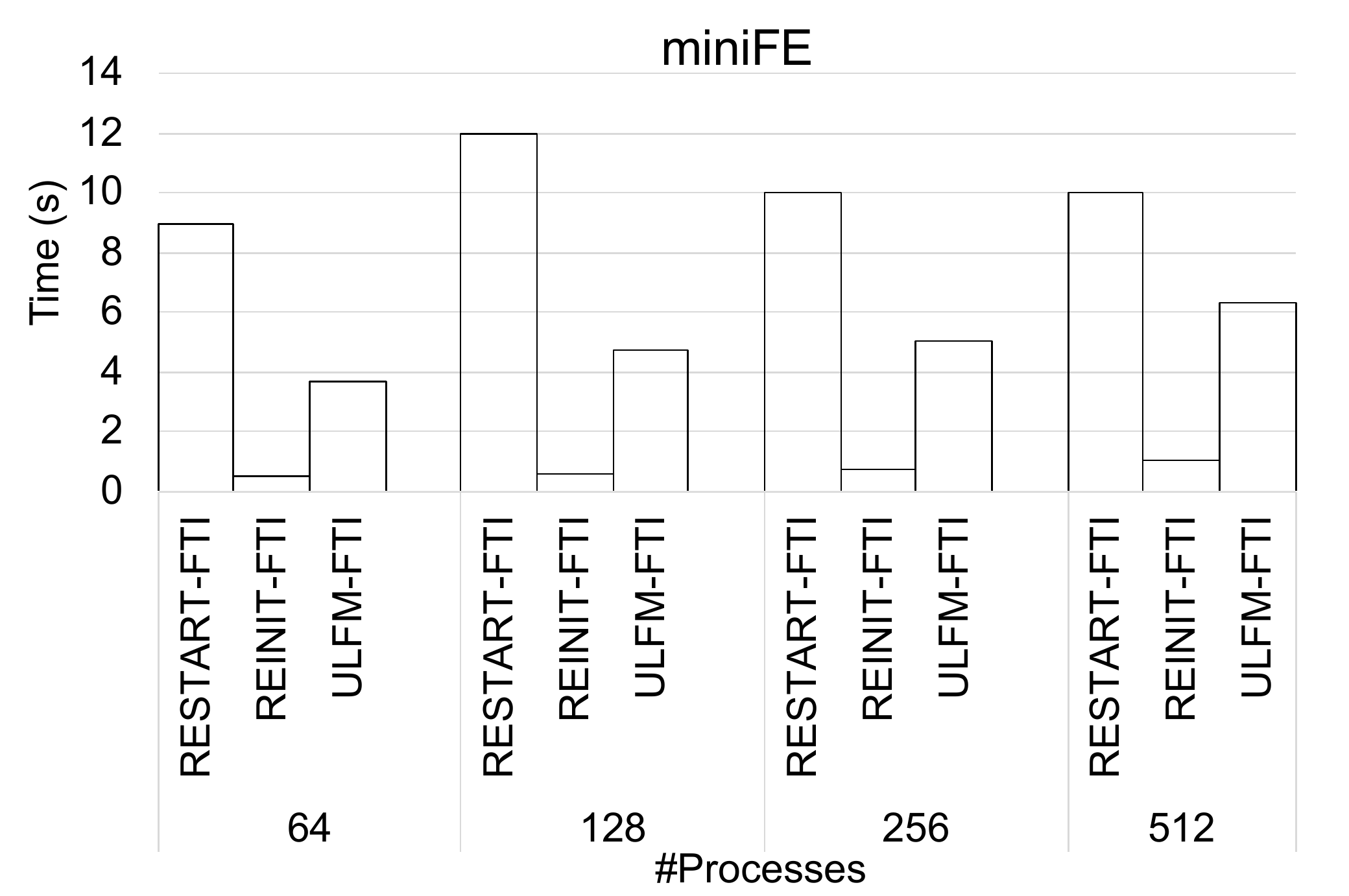}
    \caption{miniFE}
  \end{subfigure}
  \begin{subfigure}[t]{.30\textwidth}
    \includegraphics[width=1.0\textwidth]{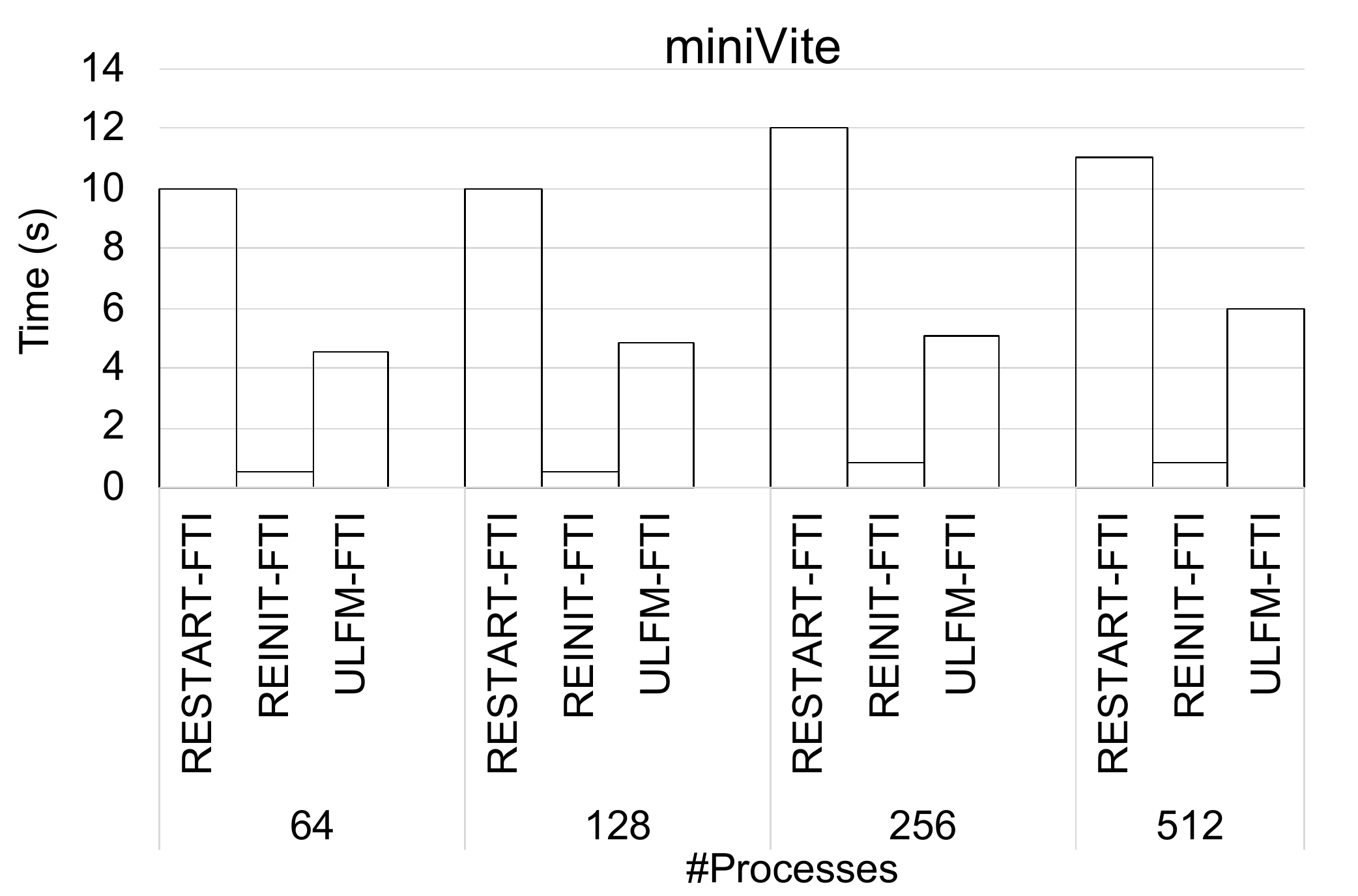}
    \caption{miniVite}
  \end{subfigure}
  \caption{Recovery time for different scaling sizes}
  \label{fig:failure_diff_scale_recov}
\end{figure*}

\begin{figure*}[!ht]
  \centering
  \begin{subfigure}[t]{.30\textwidth}
    \includegraphics[width=1.0\textwidth]{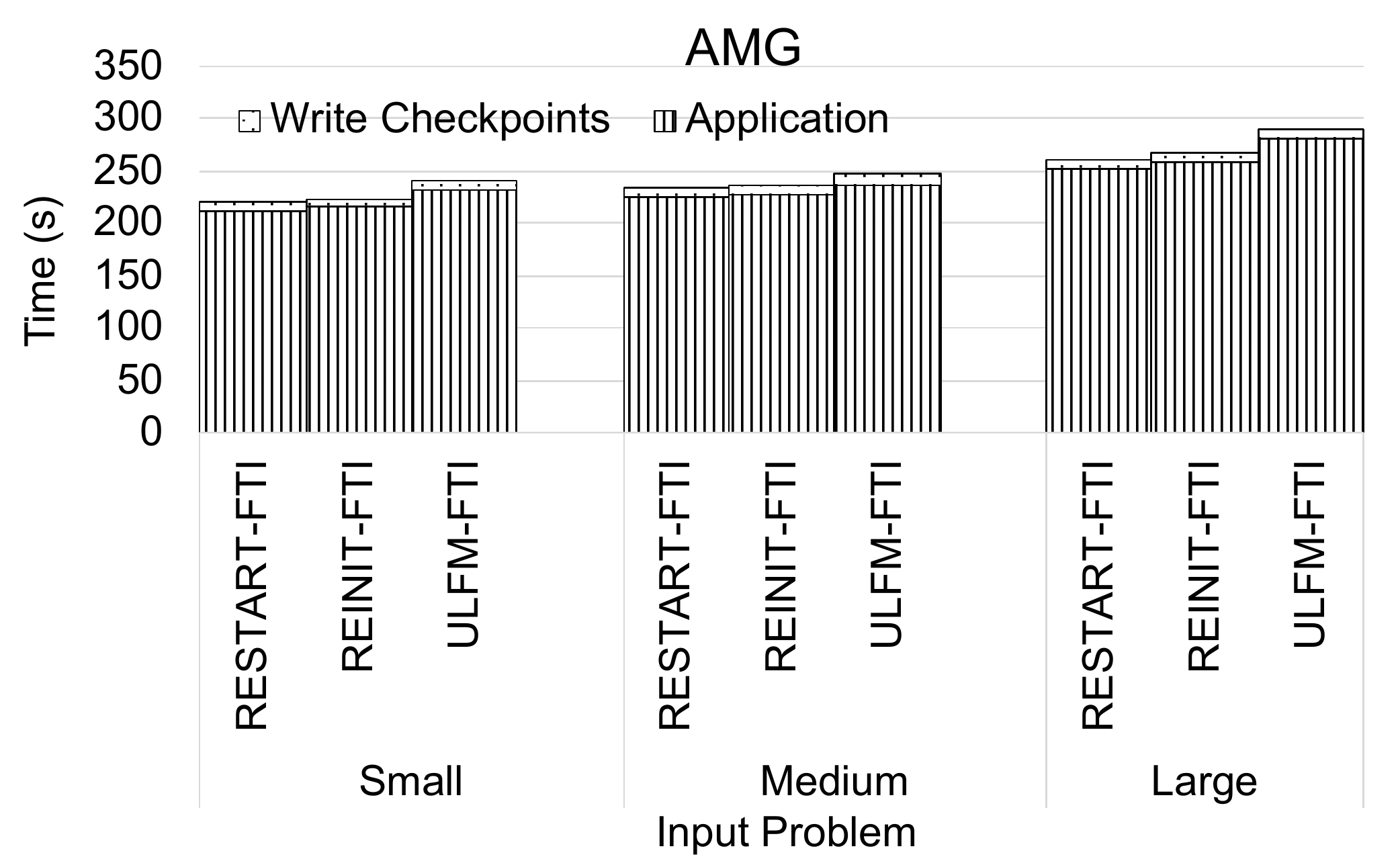}
    \caption{AMG}
  \end{subfigure}
  \begin{subfigure}[t]{.30\textwidth}
    \includegraphics[width=1.0\textwidth]{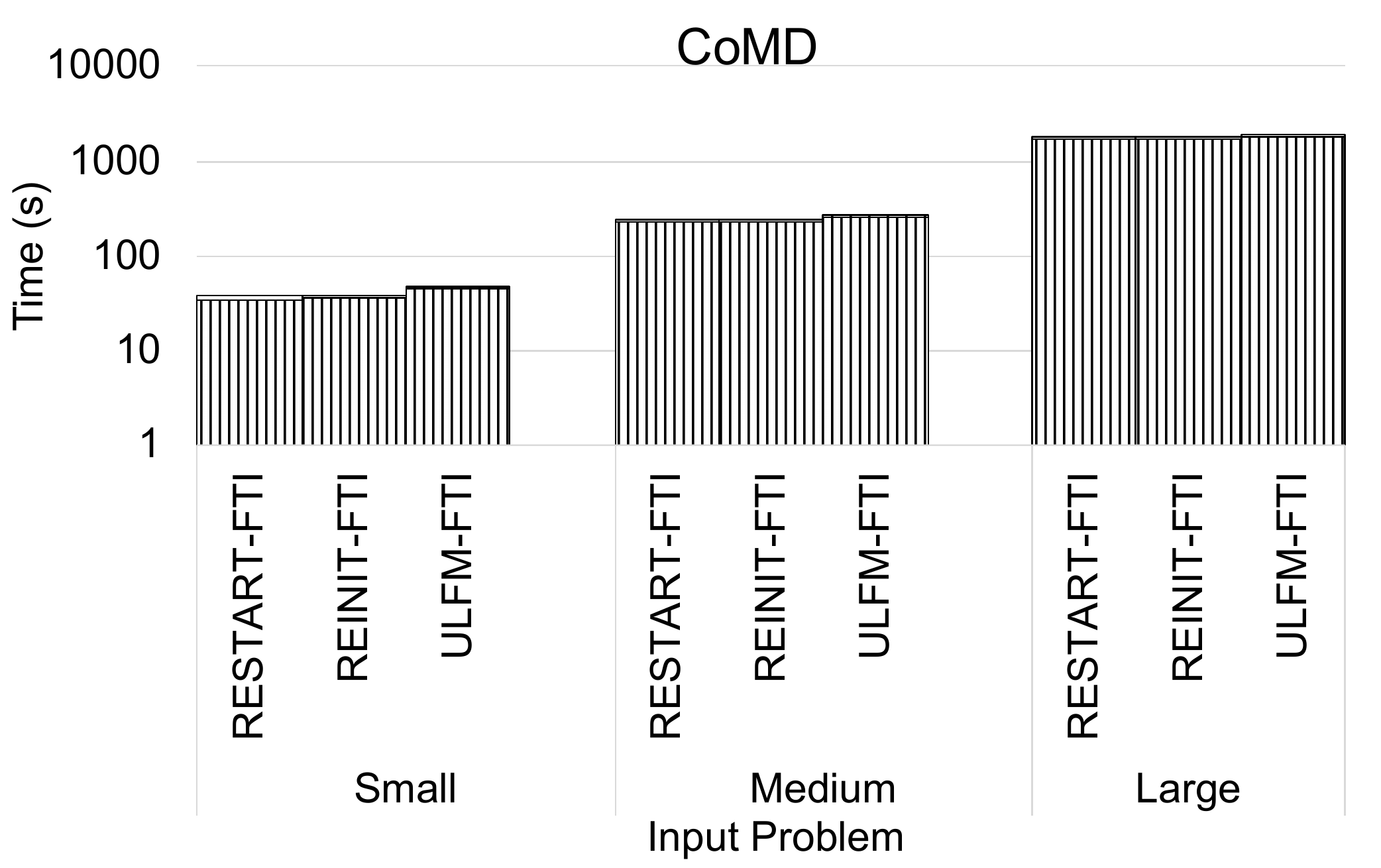}
    \caption{CoMD}
  \end{subfigure}
  \begin{subfigure}[t]{.29\textwidth}
    \includegraphics[width=1.0\textwidth]{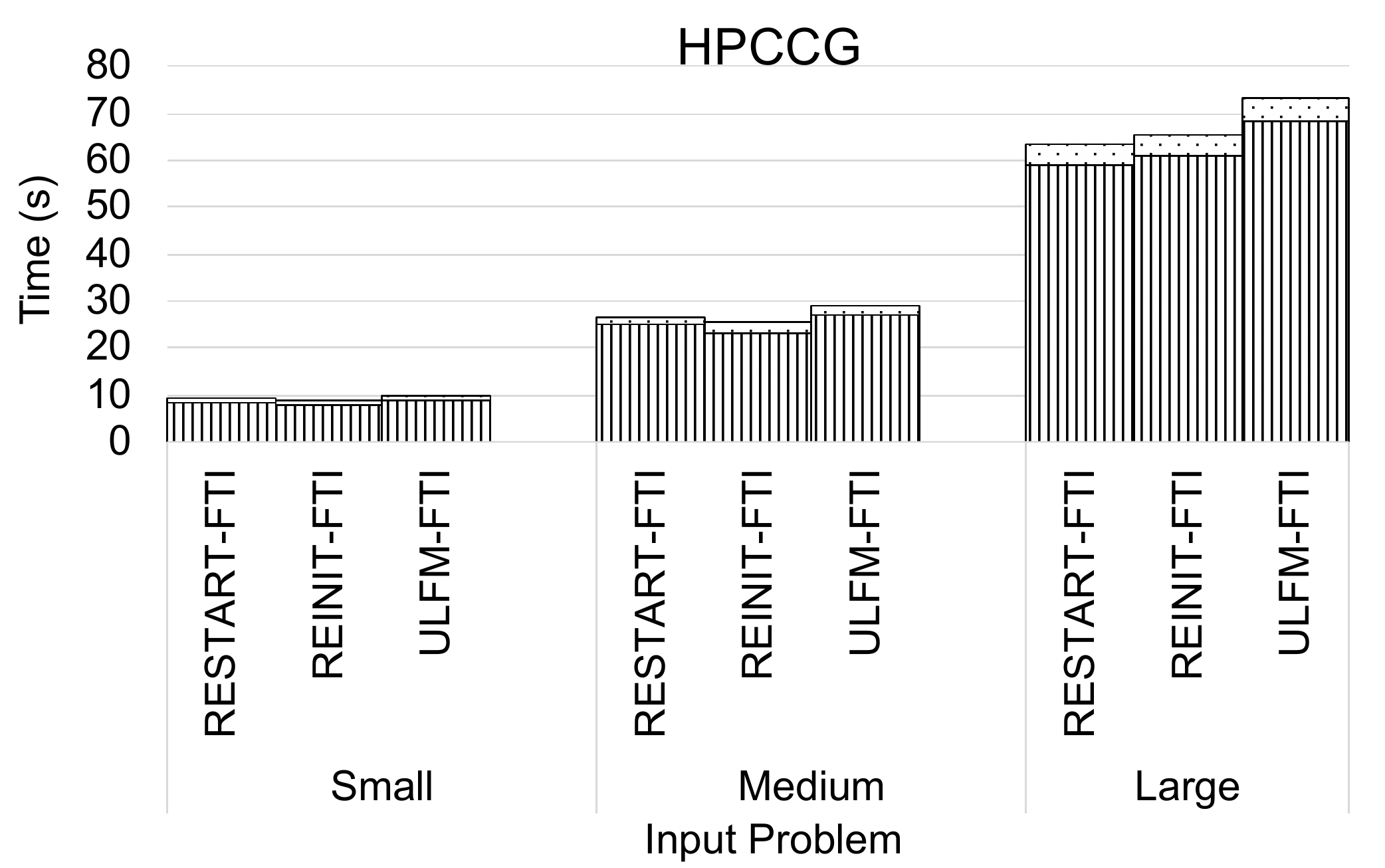}
    \caption{HPCCG}
  \end{subfigure}
    \begin{subfigure}[t]{.30\textwidth}
    \includegraphics[width=1.0\textwidth]{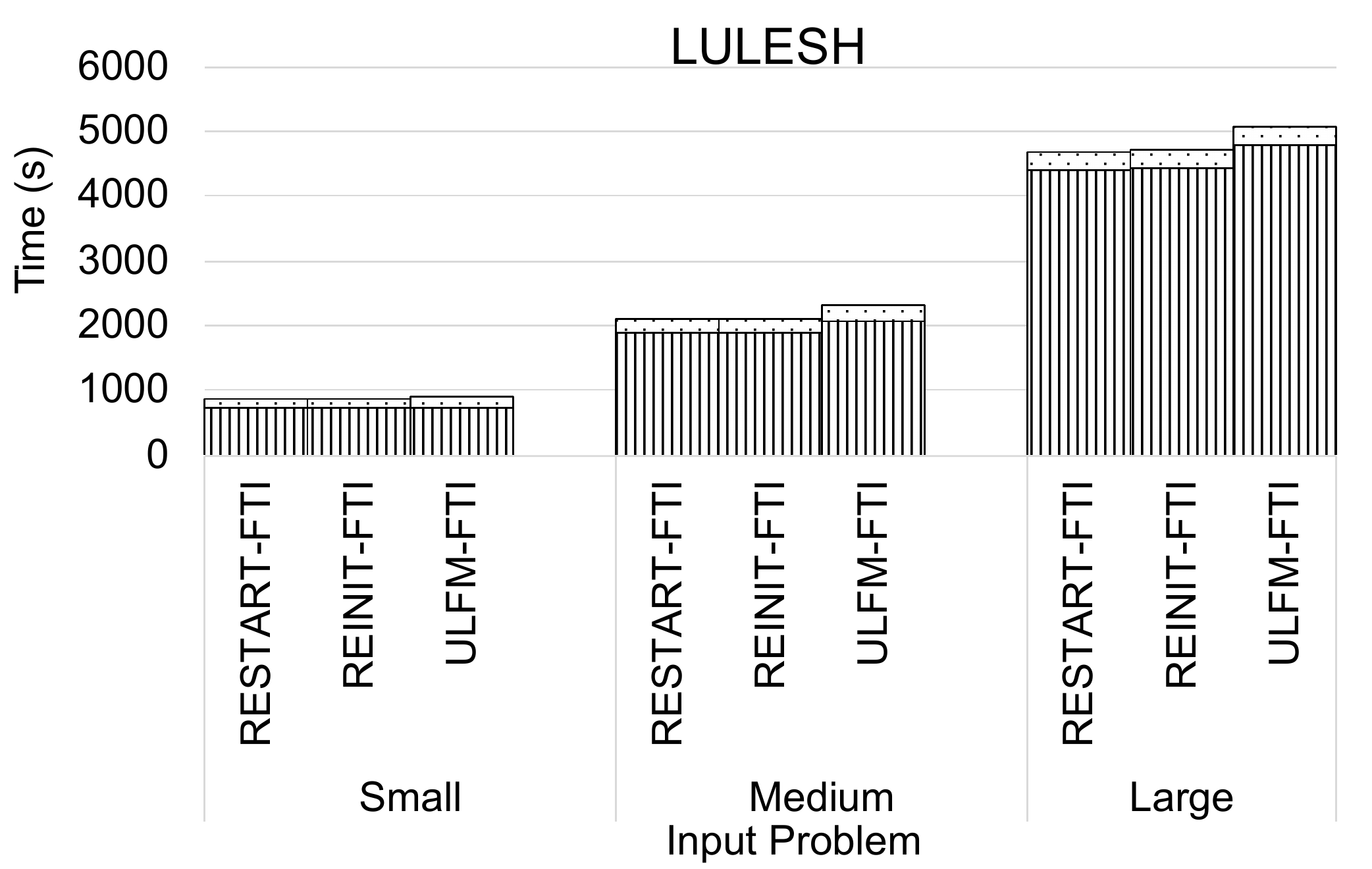}
    \caption{LULESH}
  \end{subfigure}
  \begin{subfigure}[t]{.30\textwidth}
    \includegraphics[width=1.0\textwidth]{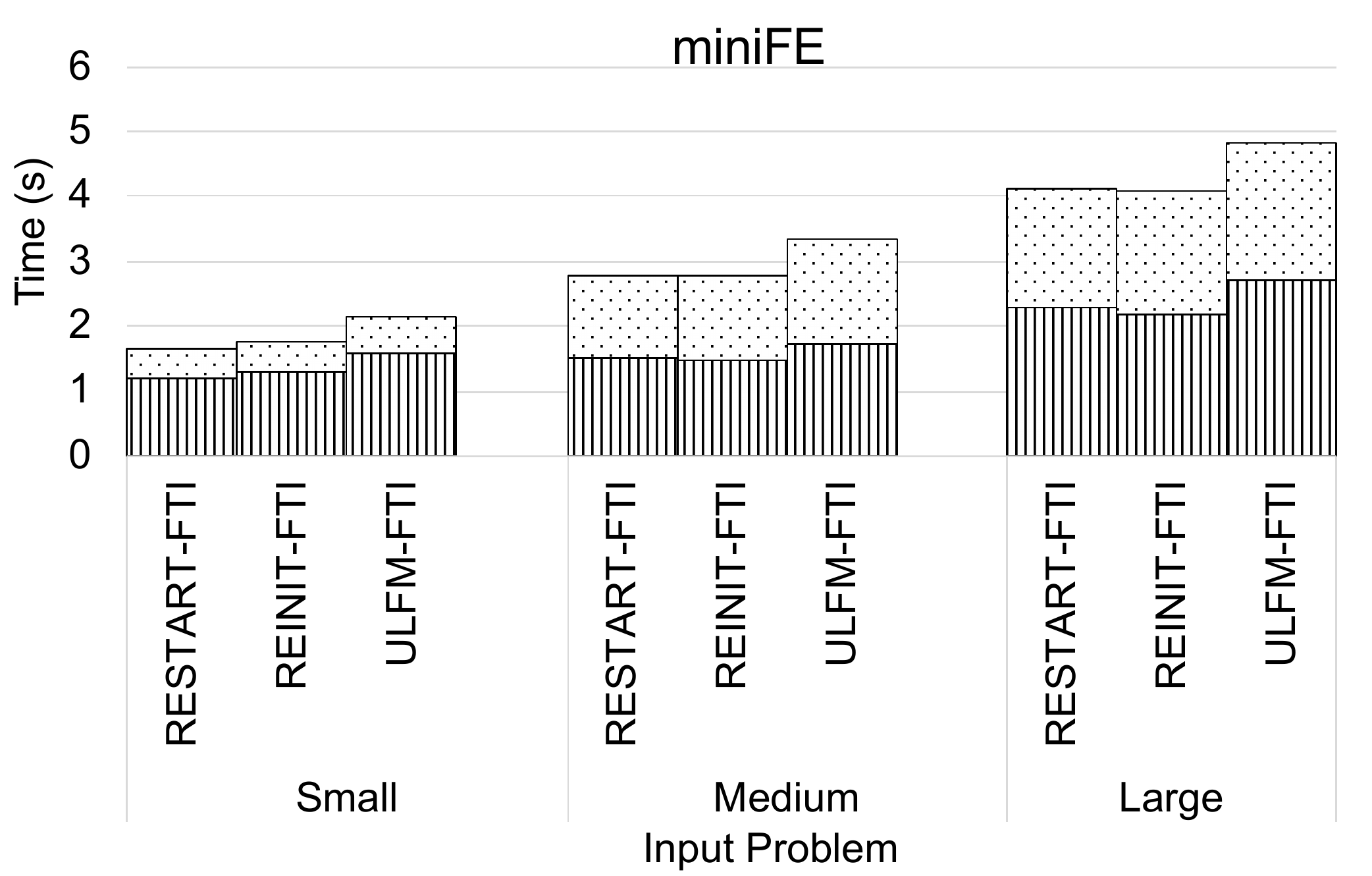}
    \caption{miniFE}
  \end{subfigure}
  \begin{subfigure}[t]{.30\textwidth}
    \includegraphics[width=1.0\textwidth]{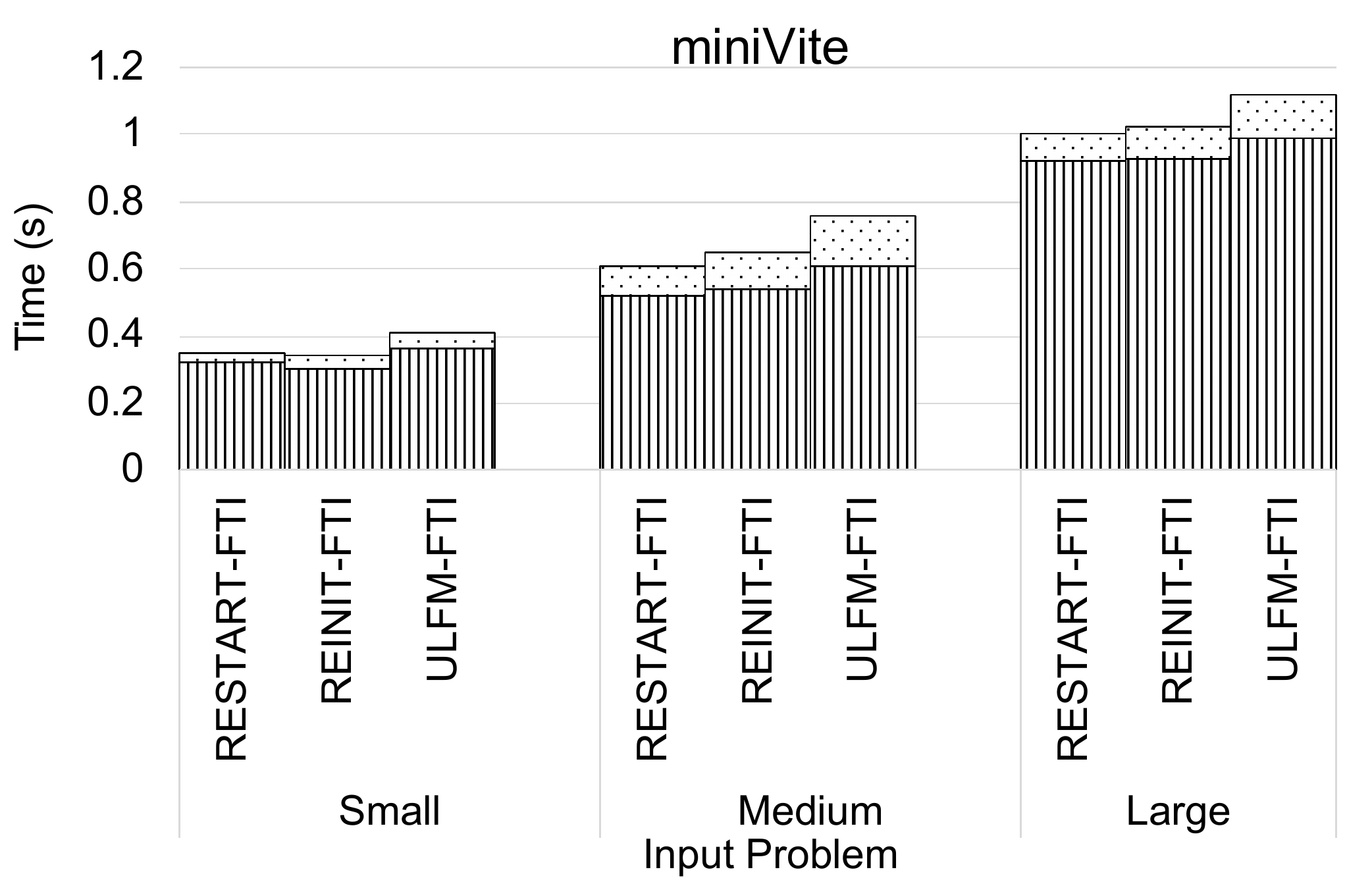}
    \caption{miniVite}
  \end{subfigure}
  \caption{Execution time breakdown in different input problem sizes with no process failures}
  \label{fig:no_failure_diff_input}
\end{figure*}

\begin{figure*}[!ht]
  \centering
  \begin{subfigure}[t]{.30\textwidth}
    \includegraphics[width=1.0\textwidth]{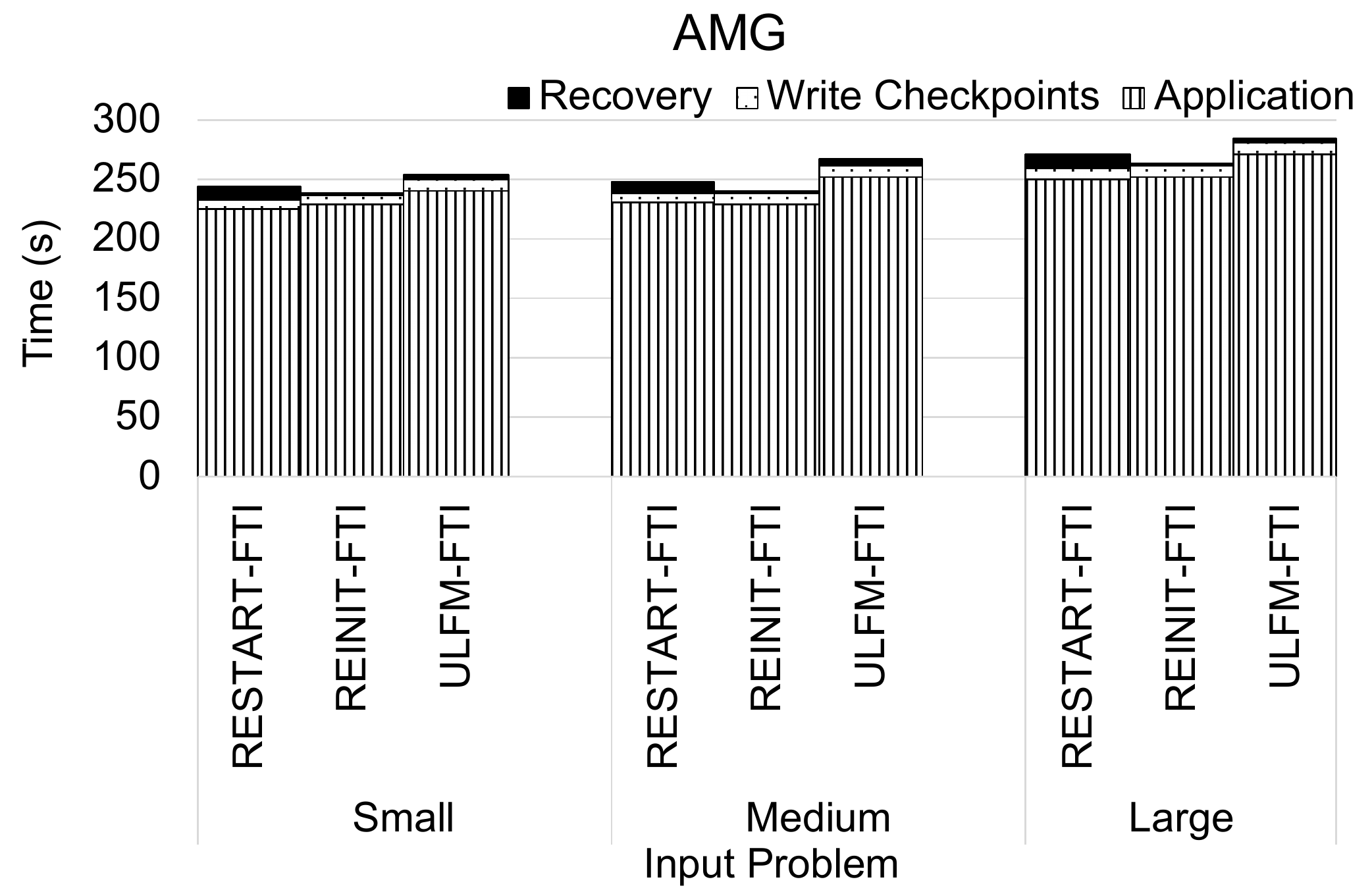}
    \caption{AMG}
  \end{subfigure}
  \begin{subfigure}[t]{.30\textwidth}
    \includegraphics[width=1.0\textwidth]{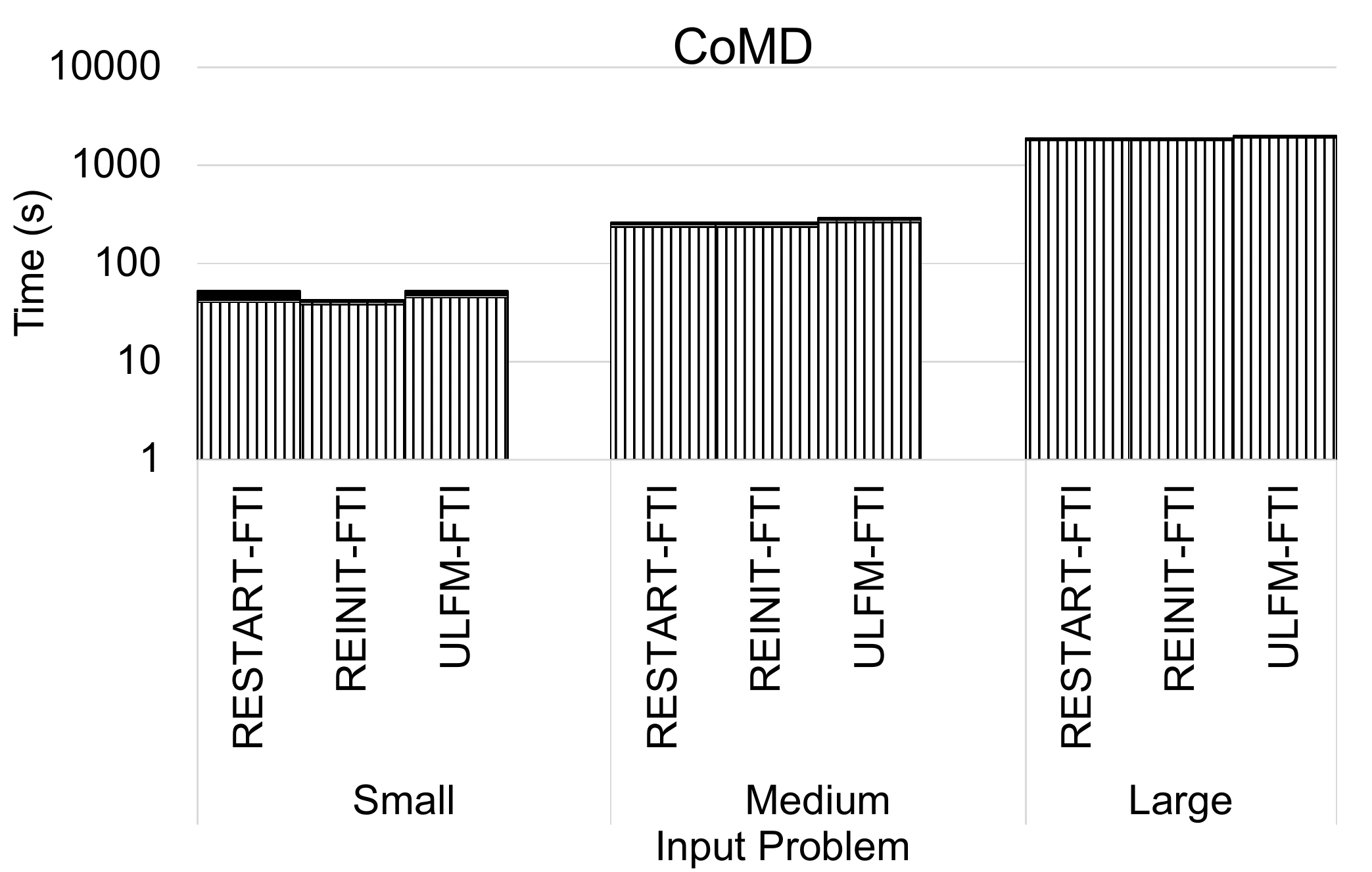}
    \caption{CoMD}
  \end{subfigure}
  \begin{subfigure}[t]{.30\textwidth}
    \includegraphics[width=1.0\textwidth]{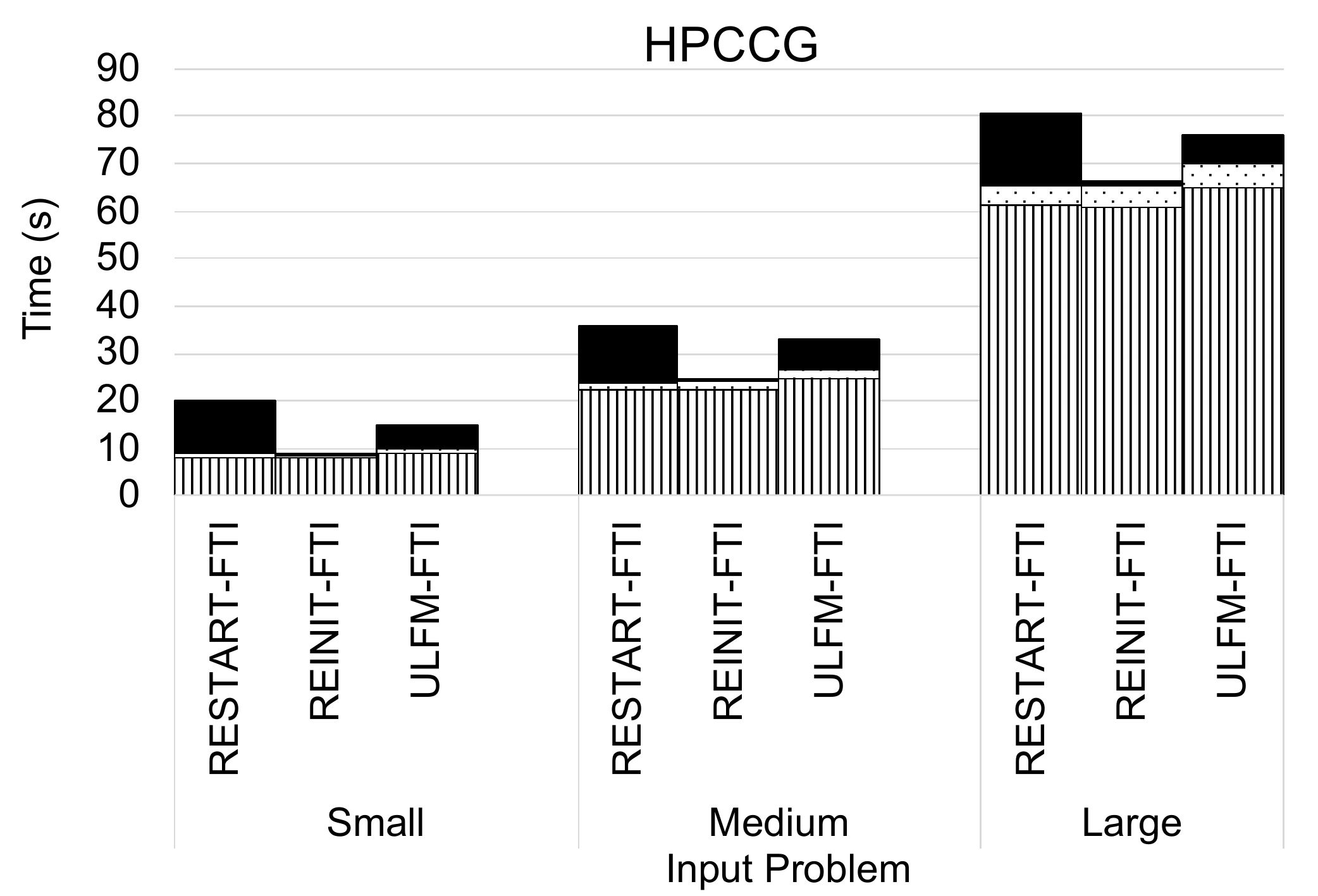}
    \caption{HPCCG}
  \end{subfigure}
    \begin{subfigure}[t]{.30\textwidth}
    \includegraphics[width=1.0\textwidth]{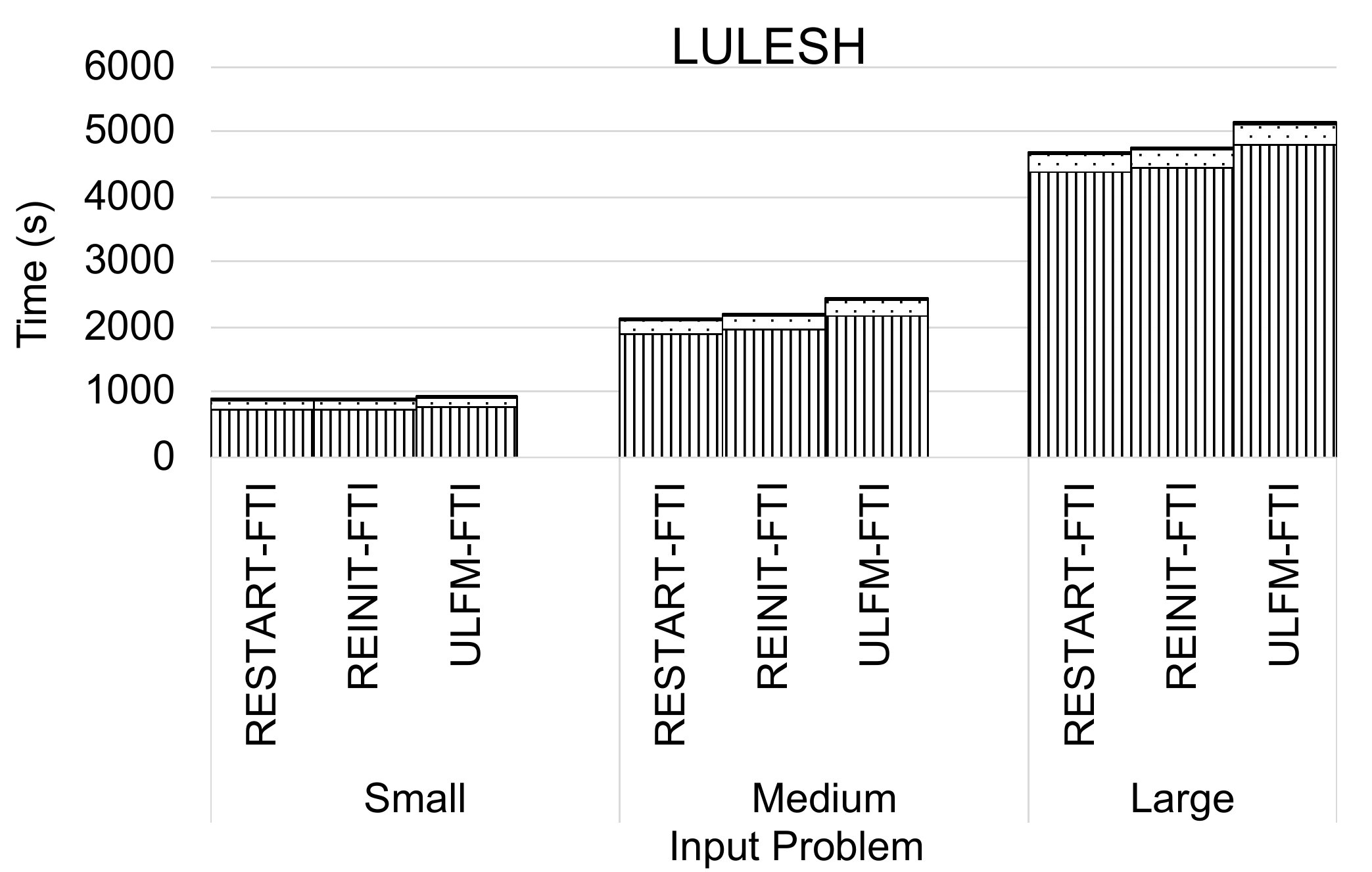}
    \caption{LULESH}
  \end{subfigure}
  \begin{subfigure}[t]{.30\textwidth}
    \includegraphics[width=1.0\textwidth]{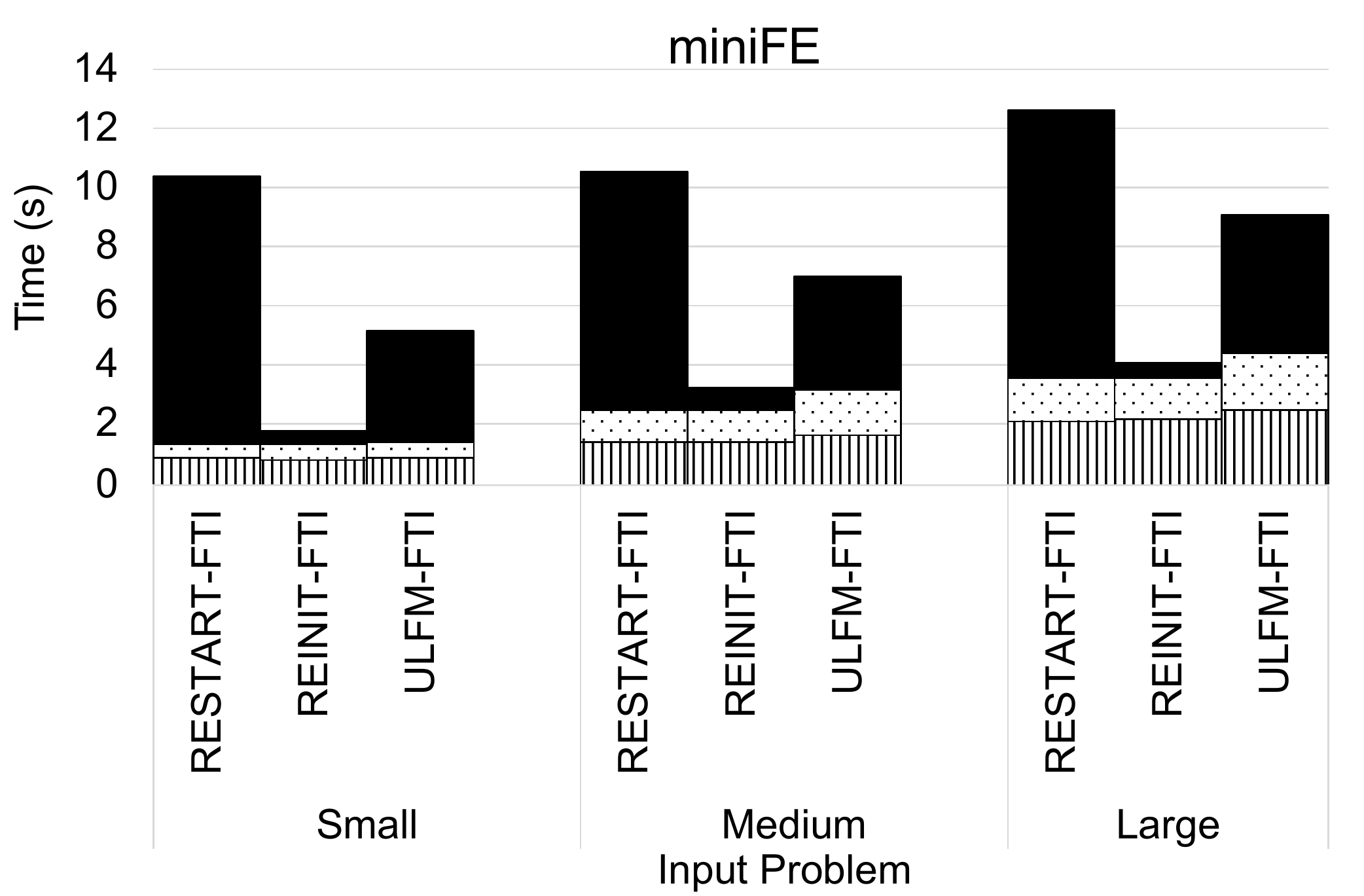}
    \caption{miniFE}
  \end{subfigure}
  \begin{subfigure}[t]{.30\textwidth}
    \includegraphics[width=1.0\textwidth]{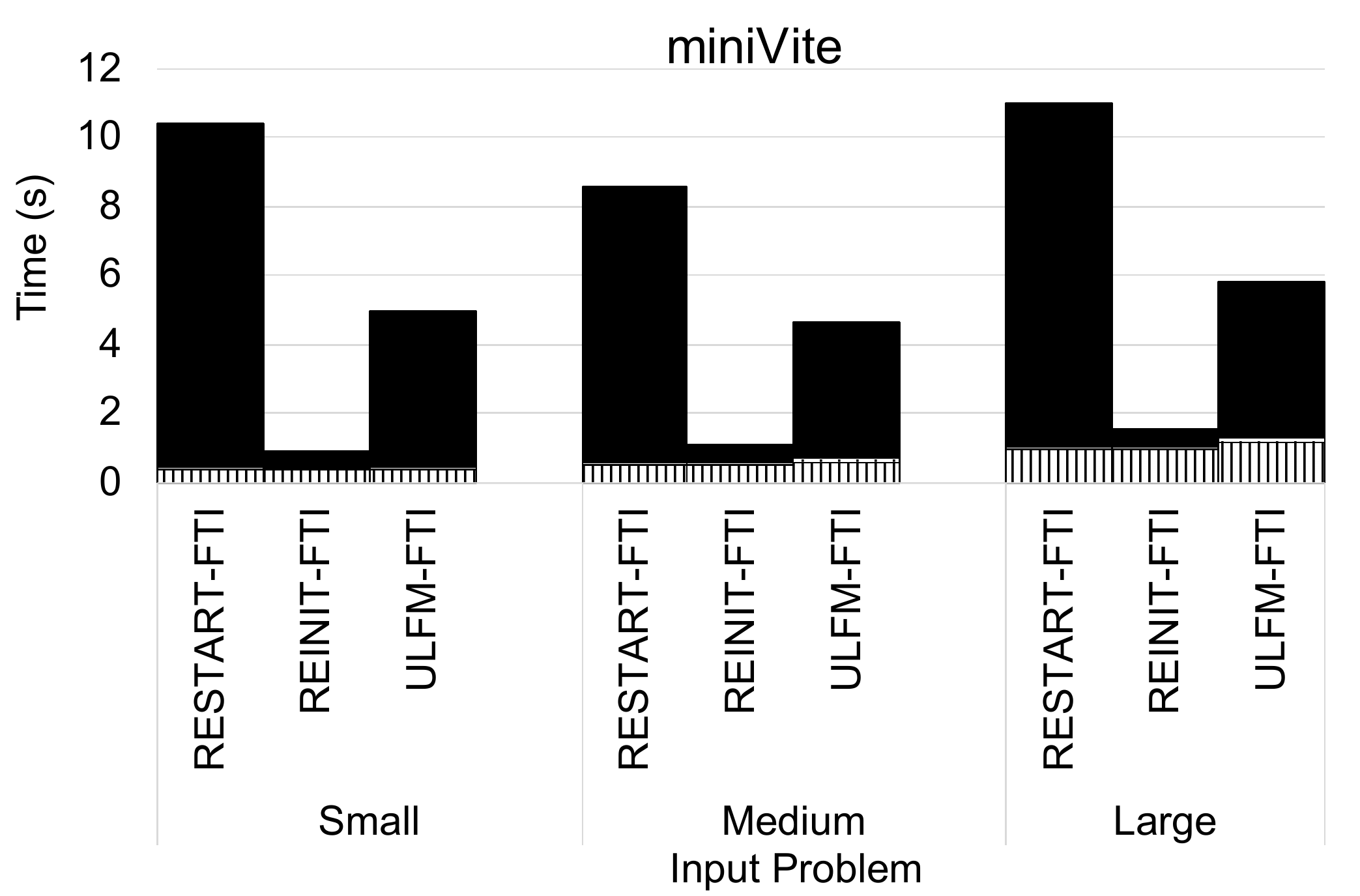}
    \caption{miniVite}
  \end{subfigure}
  \caption{Execution time breakdown recovering from a process failure in different input problem sizes}
  \label{fig:failure_diff_input}
\end{figure*}

\begin{figure*}[!ht]
  \centering
  \begin{subfigure}[t]{.30\textwidth}
    \includegraphics[width=1.0\textwidth]{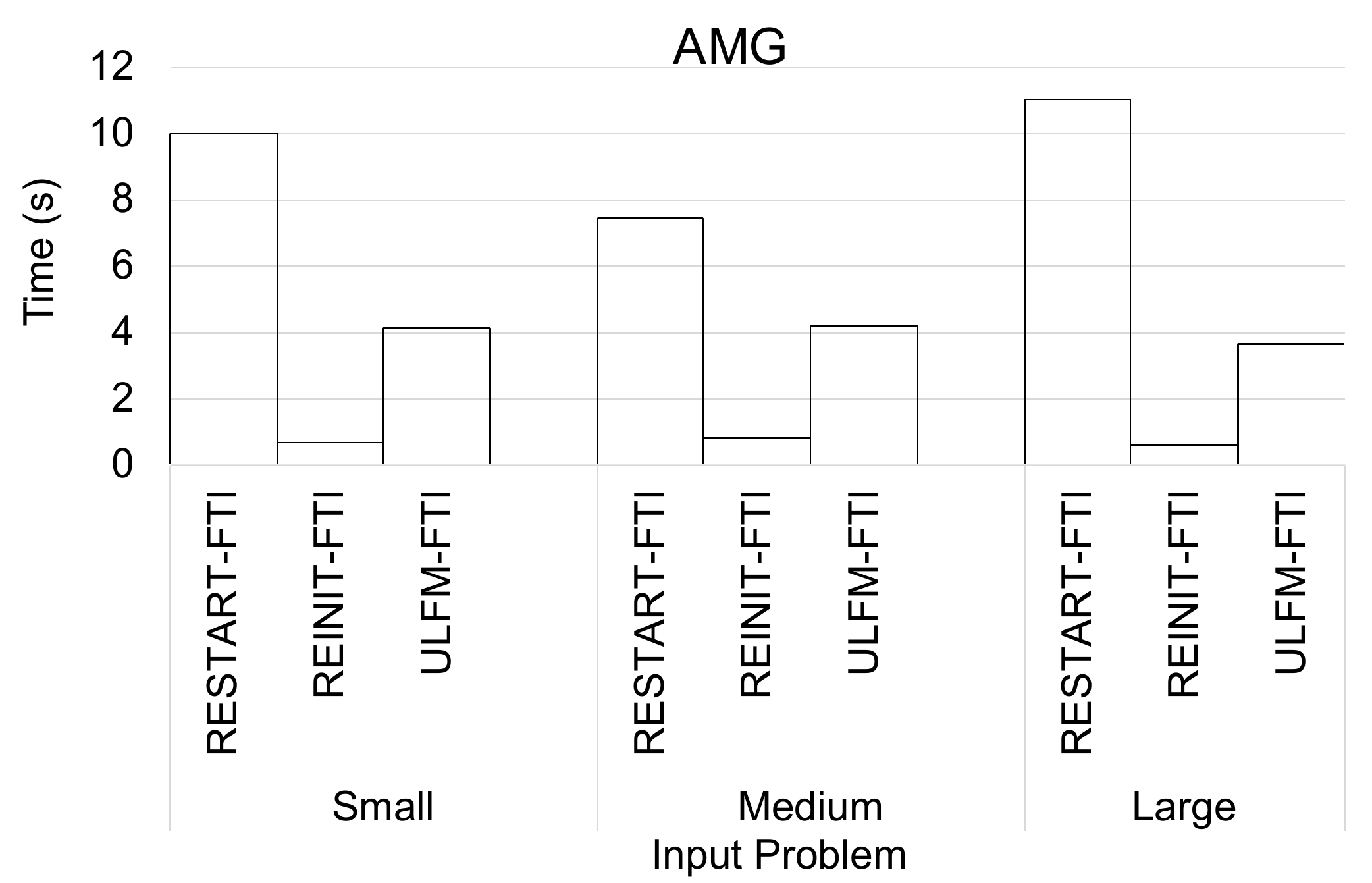}
    \caption{AMG}
  \end{subfigure}
  \begin{subfigure}[t]{.30\textwidth}
    \includegraphics[width=1.0\textwidth]{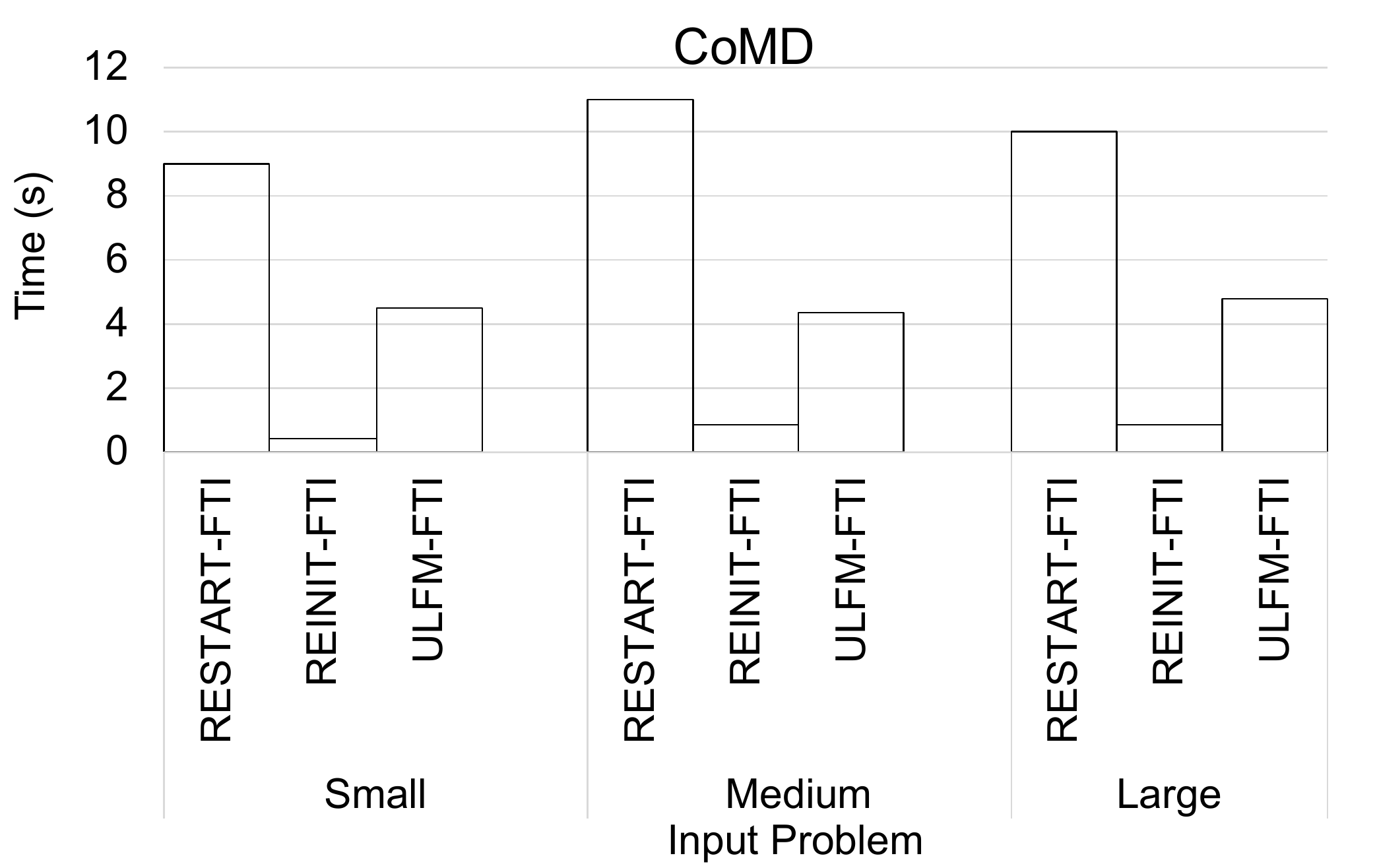}
    \caption{CoMD}
  \end{subfigure}
  \begin{subfigure}[t]{.30\textwidth}
    \includegraphics[width=1.0\textwidth]{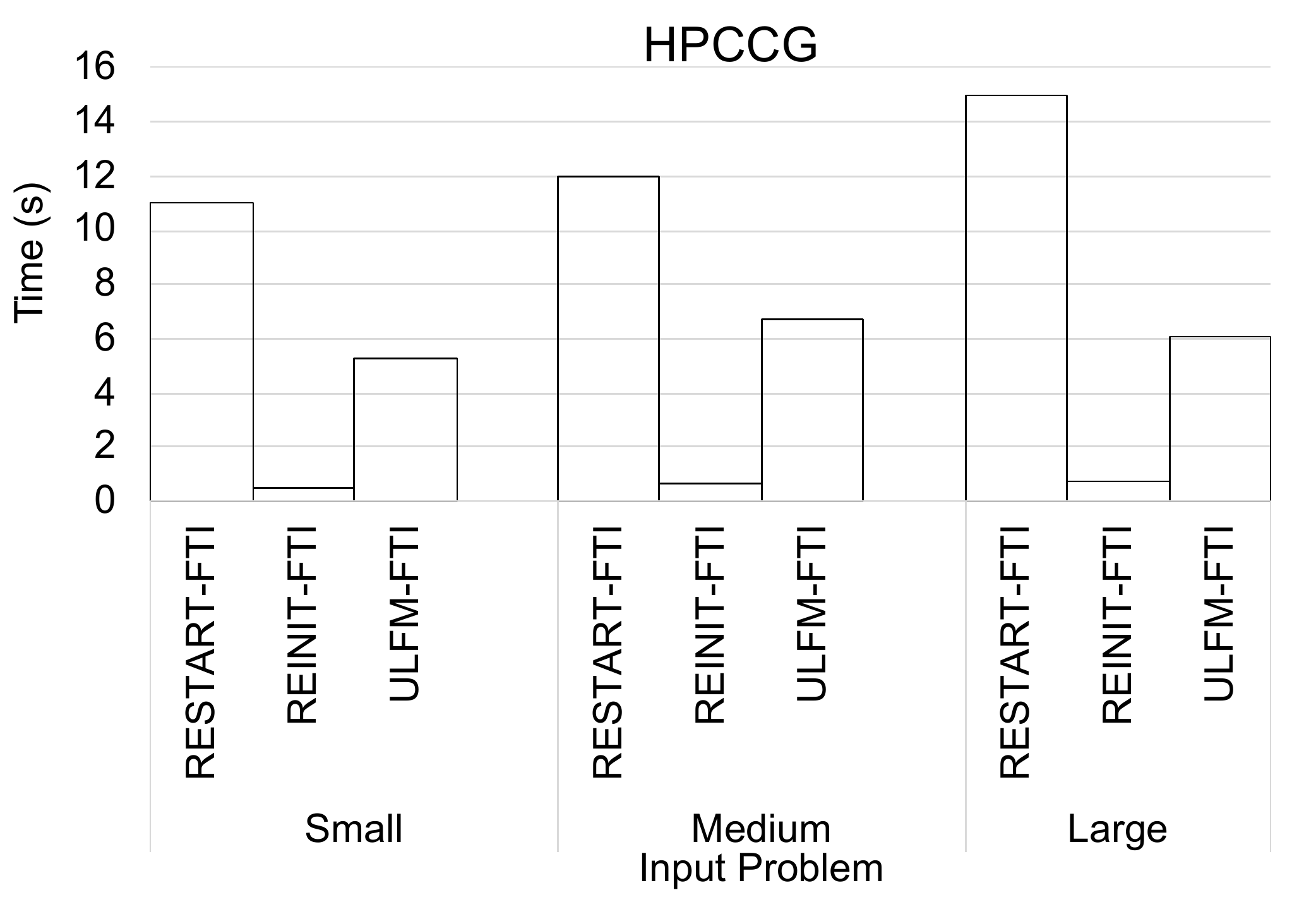}
    \caption{HPCCG}
  \end{subfigure}
    \begin{subfigure}[t]{.30\textwidth}
    \includegraphics[width=1.0\textwidth]{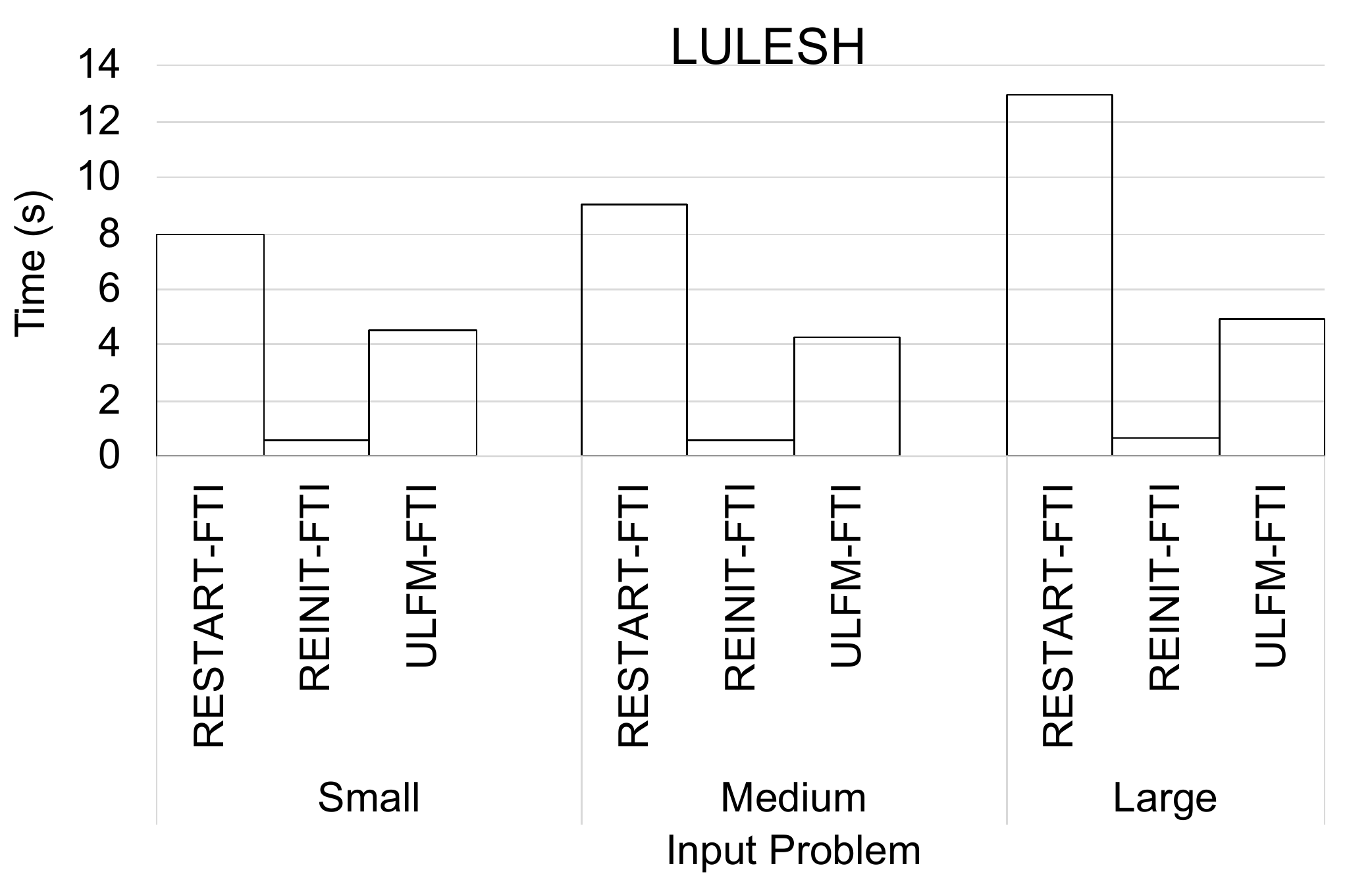}
    \caption{LULESH}
  \end{subfigure}
  \begin{subfigure}[t]{.30\textwidth}
    \includegraphics[width=1.0\textwidth]{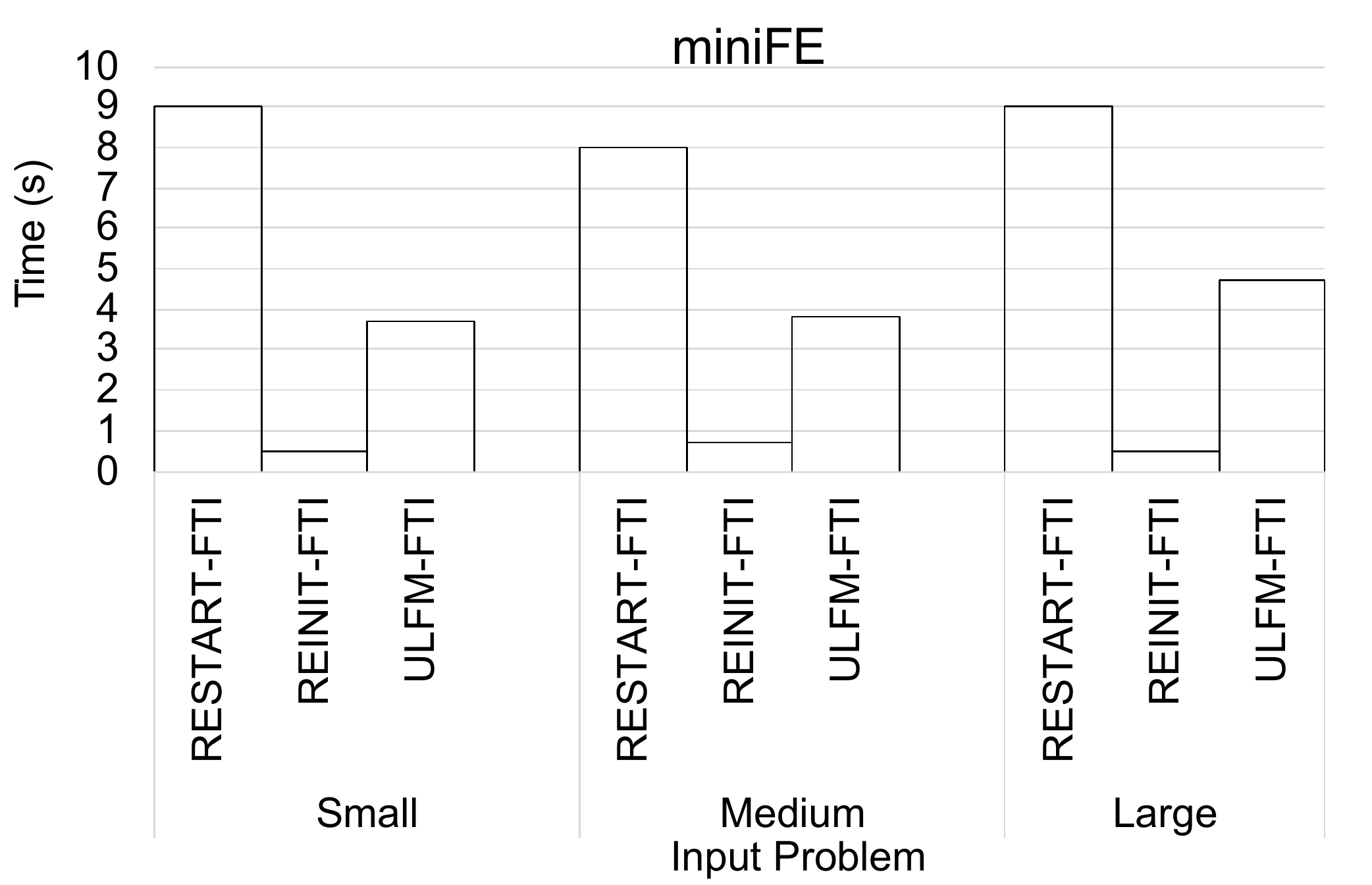}
    \caption{miniFE}
  \end{subfigure}
  \begin{subfigure}[t]{.30\textwidth}
    \includegraphics[width=1.0\textwidth]{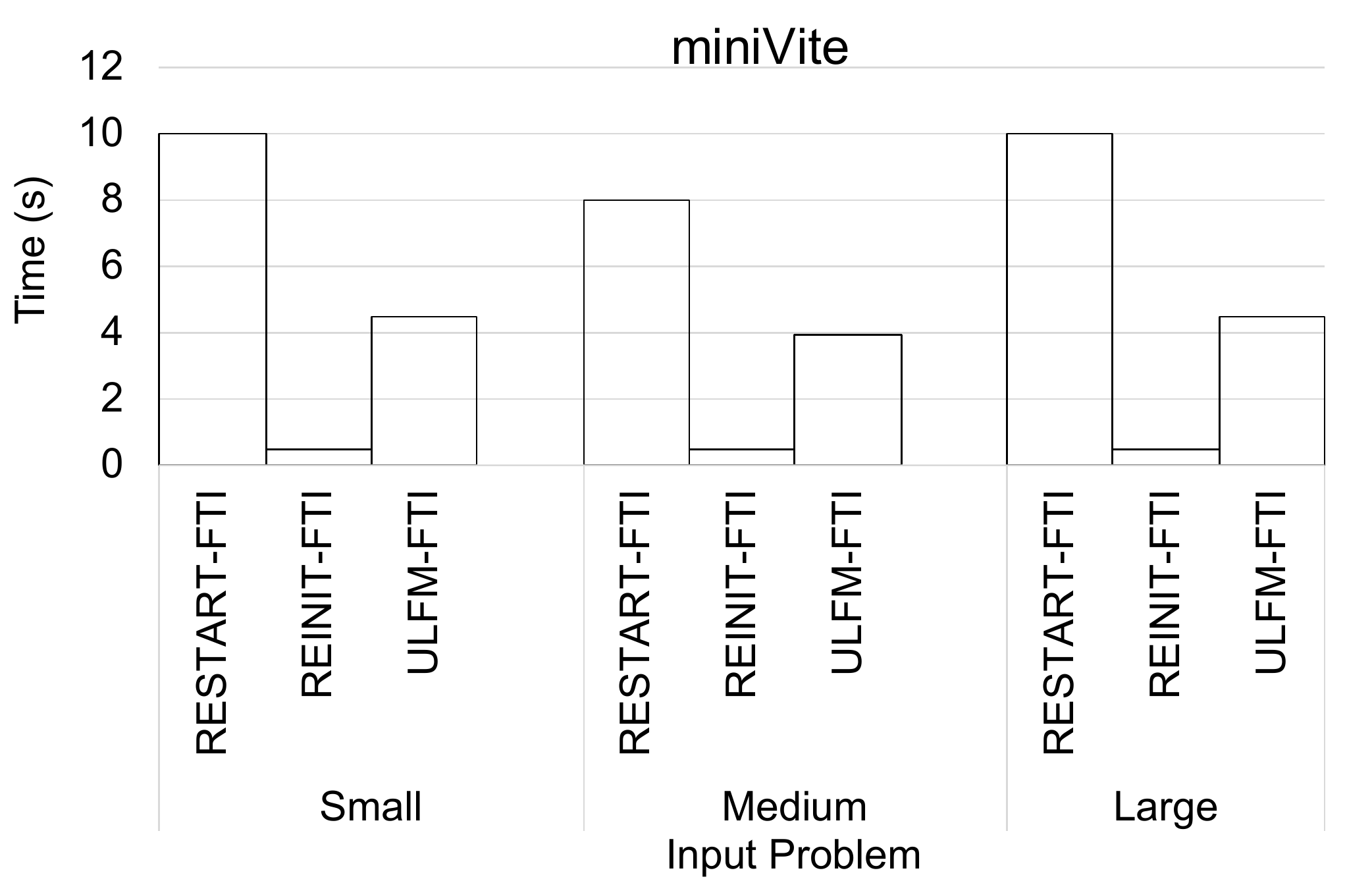}
    \caption{miniVite}
  \end{subfigure}
  \caption{Recovery time for different input problem sizes}
  \label{fig:failure_diff_input_recov}
\end{figure*}

\subsection{Artifact Description}
We run experiments on a large-scale HPC cluster having 752 nodes. Each node is equipped of two 
Intel Haswell CPUs, 28 CPU cores, 128 GB shared memory, and 8 TB local storage.

\subsection{Experimentation Setup}
This section provides details of the experimentation setup. 
We evaluate three fault tolerance designs. 
Those are \emph{FTI checkpointing with Restart (RESTART-FTI)} which means that we restart the execution, in case of a failure, for MPI recovery, 
\emph{FTI checkpointing with Reinit recovery (REINIT-FTI)}, and \emph{FTI checkpointing with ULFM recovery (ULFM-FTI)}.

For FTI checkpointing, we use its L1 mode. 
FTI L1 mode allows to store checkpoints to the local SSD or to do in-memory checkpointing.
We use the fastest approach that saves checkpoints to the local memory
using RAMFS through ``/dev/shm".
Although there are also L2, L3, and L4 modes for 
checkpointing, we do not evaluate all of them.
The efficiency comparison between the four FTI checkpointing modes has been thoroughly studied 
in the FTI paper~\cite{bautista2011fti}. 
We save checkpoints every \textbf{ten} iterations. For ULFM, we use the latest version ``ULFM v4.0.1ulfm2.1rc1" based on Open MPI 4.0.1. For Reinit~\cite{isc2020reinit}, we use its latest version based on Open MPI 4.0.0. 

We implement all the three fault tolerance designs in the MATCH workloads. 
Each design is run on three input problem sizes with the default scaling size (64 processes) with and without fault injection.
Each design is also executed on four scaling sizes (64 processes on 32 nodes, 128 processes on 
32 nodes, 256 processes on 32 nodes, and 512 processes on 32 nodes) with the default input problem size (small) with and without fault injection.
We show the experimentation configuration in Table~\ref{tab:config}.
Note that LULESH needs to run on a cube number of processes, thus runs only with 64 and 512 processes. 

For fault injection, 
we choose a \emph{random} iteration and a \emph{random} process to inject a fault.
This enables us to fairly compare the efficiency of different fault tolerance configurations. 

Notably, we run the experiment of each configuration for five times, and calculate 
the average execution time to minimize system noise. We use `-O3' for mpicc or mpicxx compilation.

\subsection{Performance Comparison on Different Scaling Sizes}

In this experiment, we run each evaluation on four scaling sizes with the default input problem size (small). 
We seek to compare the scaling efficiency of the three fault tolerance designs with and without process failures.

\textbf{Without A Failure:}
Figure~\ref{fig:no_failure_diff_scale} shows the average execution time 
with no failure. We break down the execution 
time to \textit{the application execution time} and \textit{the time to write checkpoints}. 


Overall, we can see that among the three fault tolerance designs, ULFM-FTI performs worst. RESTART-FTI and REINIT-FTI perform similar and better than ULFM-FTI. 

We first observe that FTI checkpointing scales well. The time spent 
on writing checkpoints modestly increases with more processes. This implies that there are a number of collective operations used in FTI L1 checkpointing.
The time for writing checkpoints is accounted for 13\% of the total execution time. 

Second, we observe that Reinit has no impact on application execution when there is no
failure. We use the FTI application execution time as the baseline 
for comparison because FTI is an application-level checkpointing library, whereas 
ULFM and Reinit modify the MPI runtime. We can see that the \textit{application execution time} of REINIT-FTI is very close to the \textit{application execution time} of RESTART-FTI.
However, ULFM-FTI introduces overhead to the application execution.
This overhead grows as the number of processes goes up. 
This is understandable. ULFM is implemented across MPI runtime and application levels. It can introduce memory access and communication latency to the application execution and further affect the application execution efficiency.
As reported in a ULFM paper~\cite{doi:10.1177/1094342017711505}, ULFM implements a constantly 
heartbeat mechanism for failures detection, and also amends MPI communication interfaces for failure recovery operations. 
These changes must have an impact on application execution efficiency. 
Different from ULFM, 
Reinit incurs overhead only when a failure happens because
it does not perform any background operation during application execution.


Furthermore, we observe that the times for writing checkpoints in RESTART-FTI and REINIT-FTI cases are close. This indicates that Reinit has no interference on FTI checkpointing, yet ULFM has a small impact 
on FTI checkpointing in cases such as HPCCG and miniVite. 
This is reasonable. Reinit is implemented at the MPI runtime level, 
which has a minimal impact on application-level operations, where the FTI operations run.
In contrast, ULFM does a significant amount of collective operations for a periodic heartbeat in the MPI runtime, 
which leads to background overhead.


\textbf{With A Failure:}
Figure~\ref{fig:failure_diff_scale} shows the breakdown of execution time recovering from a process failure 
on different scaling sizes.
Note that reading checkpoints only happens once in the execution, after recovery, and it is in the order of milliseconds,
which is difficult to observe, so we exclude it from the figure. 
Figure~\ref{fig:failure_diff_scale_recov} shows the MPI recovery time for different scaling sizes. 

Overall, we observe that REINIT-FTI achieves the best performance compared to RESTART-FTI and ULFM-FTI. 
There are two reasons. First, Reinit recovery does not affect application execution, including checkpoint writing. 
Second, Reinit recovery uses the least time for MPI recovery than restarting and ULFM recovery. 
Those are similar observations derived from Figure~\ref{fig:no_failure_diff_scale}.
Furthermore, we provide new observations by comparing the MPI recovery efficiency for the three fault tolerance designs. 

\textbf{ULFM recovery vs. Reinit Recovery.} We find that ULFM recovery 
time can be up to 13 times larger than Reinit recovery time, and four times larger on average. 
We can also see a trend that the ULFM recovery time increases as the number of processes grows, thus not scaling well. 
\textit{Different from ULFM, we find that Reinit recovery is independent of the number of processes.} 
 Since, ULFM enforces a variety of fault tolerance collective operations on all MPI 
processes to enable MPI global non-shrinking recovery. Even worse, ULFM implements 
these fault tolerance operations at the application level, which needs to synchronize with 
other fault tolerance operations implemented at the MPI runtime. By contrast, Reinit is 
implemented at the MPI runtime level, and Reinit requires much fewer collective operations. 

\textbf{Restart vs. Reinit recovery.} We find that the restart recovery can be up to 
22 times slower than Reinit recovery, and 16 times slower on average. This is expected. 
Redeployment of the MPI setup and allocation of resources for restarting the execution is very expensive. 
Reinit recovery repairs MPI state online.

\textbf{Restart vs. ULFM recovery.} Restart recovery is 2 to 3 times slower than ULFM recovery. 
Similarly to Reinit, ULFM recovery is online, which is much more efficient than redeployment.

\subsection{Performance Comparison on Different Input Sizes}
In this experiment, we compare the performance efficiency of the three fault tolerance designs on three input problem sizes
with the default scaling size (64 processes), with and without fault injection. 

\textbf{Without A Failure:}
Figure~\ref{fig:no_failure_diff_input} presents the results of different input problem sizes with no process failures. 
The execution time is divided into the time of writing checkpointing and pure application execution time. 
We make several observations.
Again, we use the pure application execution time of RESTART-FTI as the baseline for comparison. 

First, we can see an increment on the pure application execution time and FTI checkpointing time when running on larger input problem sizes
because the amount of data to process increases.
We can also observe the performance overhead in application execution
time in ULFM-FTI. 
This overhead increases as the input problem size grows. This indicates that ULFM is intensively 
involved in the application execution, where ULFM fault tolerance operations run a large number of 
collective MPI operations. These inefficient operations significantly affect 
the application execution, causing significant communication latency, especially when there is a 
large amount of data to process.
Different from ULFM, Reinit does not delay application execution. We can observe that 
the application execution time of REINIT-FTI is very close to the execution time of RESTART-FTI. 
This is expected as Reinit is implemented in the MPI runtime. Also, Reinit uses much fewer collective operations than ULFM does. 


\textbf{With A Failure:}
Figure~\ref{fig:failure_diff_input} shows the execution time breakdown in a process failure in different input problem sizes. 
Note that we omit the time of reading checkpoints because it is in the order of milliseconds.
Also, Figure~\ref{fig:failure_diff_input_recov} shows the recovery time for different input problem sizes. 

The same observations from Figure~\ref{fig:no_failure_diff_input} and the scaling experiments still hold. The new observation is that the recovery time of either ULFM or Reinit negligibly 
changes for different input problem sizes. This is an interesting finding but follows from their operation. 
When a failure occurs, ULFM starts collecting messages among daemons and processes in the background, while the program execution terminates. ULFM recovery dominates the execution. 
Reinit is fully implemented in the MPI runtime, which is even more difficult to be affected. 
We conclude that ULFM and Reinit recovery are \emph{independent} of the input problem size. 

\textbf{Conclusion.} \textit{(1) ULFM delays application execution, whereas Reinit has a negligible impact on the application execution; (2) ULFM affects the performance of FTI checkpointing, whereas Reinit has a negligible effect on it; (3) Reinit performs better than ULFM both when there is no failure and when there is a failure for MPI global, backward, non-shrinking recovery; (4) REINIT-FTI is the most efficient design within the three fault tolerance designs for MPI global, backward, non-shrinking recovery.} 

\subsection{Use of MATCH}

MATCH can help HPC programmers aiming at MPI fault tolerance in three ways. 
(1) We provide hands-on instructions of implementing ULFM with FTI, Reinit with FTI, and FTI with Restart on representative HPC proxy applications. MATCH is open-source. 
Programmers can learn with less effort on how to implement the three fault tolerance designs to an HPC application through the MATCH code. 
(2) We provide a data dependency analysis tool to identify data objects for checkpointing. 
Those data objects are the minimal data objects needed to guarantee the application execution correctness after restoring the application state. 
This is especially useful for applications with many data objects that need to be checkpointed.
(3) MATCH can also be a foundation for future MPI fault tolerance designs. 
Programmers can develop new MPI fault tolerance designs on top of the three fault tolerance designs. 
For example, the ULFM global non-shrinking recovery can be replaced with the ULFM local forward recovery; the FTI checkpointing can be replaced with the SCR checkpointing~\cite{6494566}. 
This is also part of our future work. 
Lastly, we encourage programmers to add new HPC applications and new MPI fault tolerance designs to MATCH. 

\section{Related Work}
\noindent\textbf{Data Recovery.}
Checkpointing~\cite{hargrove2006berkeley,sankaran2005lam,adam2019checkpoint,subasi2018unified,wang2018fault,cao2016system,adam2018transparent,kohl2019scalable}
is the commonly used approach to restart an MPI application when a failure occurs. 
Hargrove et al.~\cite{hargrove2006berkeley} develop a
system-level checkpointing library--the Berkeley Lab Checkpoint/Restart (BLCR) library--to run checkpointing at system-level using the Linux kernel.
Furthermore, Adam et al.~\cite{adam2019checkpoint}, SCR~\cite{6494566}, and FTI~\cite{6114441} propose multi-level checkpointing
aiming to significantly advance checkpointing efficiency.
CRAFT~\cite{shahzad2018craft} provides a fault tolerance framework
that integrates checkpointing to ULFM shrinking and non-shrinking recovery.
In this work, we choose FTI for checkpointing for data recovery because of the high efficiency and its extenstive documentation.
We plan to integrate and evaluate more checkpointing mechanisms in addition to FTI in future work. 
Furthermore, different to existing work, we also provide a data dependency analytic tool to aid programmers in identifying data objects for checkpointing.



\noindent\textbf{MPI Recovery.}
ULFM~\cite{bland2013post,bland2015lessons} is a leading MPI
recovery framework in progress with the MPI Fault 
Tolerance Working Group. 
ULFM provides new MPI interfaces to remove failed processes and add new processes to communicators, and to perform resilient agreement between processes. 
ULFM requests programmers to implement shrinking or non-shrinking recovery using these interfaces. 
ULFM provides flexibility to programmers, but there is significant learning effort before programmers can correctly use ULFM interfaces to  
implement ULFM recovery. 
A large body of work~\cite{katti2015scalable,herault2015practical,bouteiller2015plan,Laguna:2014:EUF:2642769.2642775,losada2017resilient} has explored and extended the applicability of ULFM. 
Teranishi et al.~\cite{teranishi2014toward} replace failed processes with spare processes to accelerate ULFM recovery.
Bosilca et al.~\cite{doi:10.1177/1094342017711505} and Katti et al.~\cite{katti2018epidemic} propose a series of efficient fault detection mechanisms for ULFM.  

Reinit~\cite{isc2020reinit,Laguna:2014} is a more efficient solution for MPI global recovery.
Reinit hides the MPI process recovery from programmers by implementing it in the MPI runtime. Reinit provides a simple interface to programmers to define
a global restart point, in the form of a resilient target function.
The early versions~\cite{doi:10.1002/cpe.4863,doi:10.1177/1094342015623623,Laguna:2014,SULTANA20191} of Reinit have limited usage because they require hard-to-deploy changes to job schedulers. 
Most recently, Georgakoudis et al.~\cite{isc2020reinit}  propose a new design and implementation of Reinit into the Open MPI runtime.

Later, researchers realize the efficiency of combining checkpointing and MPI recovery
for higher efficiency of MPI fault tolerance. For example, FENIX~\cite{Gamell2014Fenix} and CRAFT~\cite{shahzad2018craft} both design and develop a checkpointing
interface that supports data recovery for ULFM shrinking and non-shrinking recovery. 
However, developers must explicitly manage and redistribute the restored data among survivor processes in
case of a non-shrinking recovery. This can easily cause load imbalance problems.
Also, they only evaluate their frameworks on two applications, and do not compare their fault tolerance frameworks to 
other fault tolerance designs. 
In conclusion, there is no existing work that either benchmarks the design and implementation of
MPI fault tolerance, or compares the performance efficiency of 
different fault tolerance designs.
Different from FENIX and CRAFT, we evaluate and comprehensively compare fault tolerance designs that combine FTI checkpointing and MPI recovery frameworks (ULFM and Reinit) on a collection of HPC proxy applications. 

\noindent\textbf{MPI Fault Tolerance Benchmarking.}
There have been many benchmark suites~\cite{bull2009microbenchmark,luszczek2006hpc,agarwal2014design} developed for MPI performance modeling. 
SKaMPI~\cite{reussner1998skampi} is an early benchmark suite that evaluates different implementations of MPI. 
Bureddy et al.~\cite{bureddy2012omb} develop a benchmark suite to evaluate point-to-point, multi-pair, and collective MPI communication on GPU clusters.
However, there is no MPI benchmark suite that focuses on fault tolerance and evaluates fault tolerance designs in MPI. 
This paper proposes a benchmark suite MATCH for benchmarking MPI fault tolerance.


\section{Conclusions}
MPI fault tolerance is becoming an increasingly critical problem as supercomputers continue to grow in size. 
We have designed and implemented a benchmark suite, called MATCH, to evaluate MPI fault tolerance approaches. Our benchmark suite has six representative HPC proxy applications selected from existing, flagship HPC benchmark suites. We comprehensively evaluate and compare the performance efficiency of three different fault-tolerance designs, implemented on the selected applications. The evaluation results reveal that FTI checkpointing with Reinit recovery is the most efficient fault tolerance design of those three. Our analysis and finding provide significant insight to HPC developers on MPI fault tolerance.

\section{Acknowledgments}
This work was performed under the auspices of the U.S. Department of
Energy by Lawrence Livermore National Laboratory under Contract
DE-AC52-07NA27344 (LLNL-CONF-812453). 
This research was supported by the
Exascale Computing Project (17-SC-20-SC), a collaborative effort of the
U.S. Department of Energy Office of Science and the National Nuclear
Security Administration.
This research is partially supported by U.S. National Science Foundation (CNS-1617967, CCF-1553645 and CCF-1718194).
We wish to thank the Trusted CI, the NSF Cybersecurity Center of Excellence, NSF Grant Number ACI-1920430, for assisting our project with cybersecurity challenges.

\bibliographystyle{IEEEtran}
\bibliography{li,li2,li3,li_sc16_resilience_modeling,paper,paper2}

\begin{thebibliography}{10}
\providecommand{\url}[1]{#1}
\csname url@samestyle\endcsname
\providecommand{\newblock}{\relax}
\providecommand{\bibinfo}[2]{#2}
\providecommand{\BIBentrySTDinterwordspacing}{\spaceskip=0pt\relax}
\providecommand{\BIBentryALTinterwordstretchfactor}{4}
\providecommand{\BIBentryALTinterwordspacing}{\spaceskip=\fontdimen2\font plus
\BIBentryALTinterwordstretchfactor\fontdimen3\font minus
  \fontdimen4\font\relax}
\providecommand{\BIBforeignlanguage}[2]{{%
\expandafter\ifx\csname l@#1\endcsname\relax
\typeout{** WARNING: IEEEtran.bst: No hyphenation pattern has been}%
\typeout{** loaded for the language `#1'. Using the pattern for}%
\typeout{** the default language instead.}%
\else
\language=\csname l@#1\endcsname
\fi
#2}}
\providecommand{\BIBdecl}{\relax}
\BIBdecl

\bibitem{dongarra2013emerging}
J.~Dongarra, ``Emerging heterogeneous technologies for high performance
  computing,'' in \emph{International Heterogeneity in Computing Workshop},
  2013.

\bibitem{di2014lessons}
C.~Di~Martino, Z.~Kalbarczyk, R.~K. Iyer, F.~Baccanico, J.~Fullop, and
  W.~Kramer, ``Lessons learned from the analysis of system failures at
  petascale: The case of blue waters,'' in \emph{IEEE/IFIP International
  Conference on Dependable Systems and Networks}.\hskip 1em plus 0.5em minus
  0.4em\relax IEEE, 2014.

\bibitem{ghi2016lessons}
S.~Ghiasvand, F.~M. Ciorba, R.~Tsch{\"u}ter, and W.~E. Nagel, ``Lessons learned
  from spatial and temporal correlation of node failures in high performance
  computers,'' in \emph{Euromicro International Conference on Parallel,
  Distributed, and Network-Based Processing}.\hskip 1em plus 0.5em minus
  0.4em\relax IEEE, 2016.

\bibitem{sc18:fliptracker}
L.~Guo, D.~Li, I.~Laguna, and M.~Schulz, ``Fliptracker: Understanding natural
  error resilience in hpc applications,'' in \emph{{SC}}, 2018.

\bibitem{ipdps19:guo}
L.~Guo and D.~Li, ``{MOARD: Modeling Application Resilience to Transient Faults
  on Data Objects},'' in \emph{{International Parallel and Distributed
  Processing Symposium}}, 2019.

\bibitem{katti2015scalable}
A.~Katti, G.~Di~Fatta, T.~Naughton, and C.~Engelmann, ``Scalable and fault
  tolerant failure detection and consensus,'' in \emph{Proceedings of the 22nd
  European MPI Users' Group Meeting}, 2015, p.~13.

\bibitem{herault2015practical}
T.~Herault, A.~Bouteiller, G.~Bosilca, M.~Gamell, K.~Teranishi, M.~Parashar,
  and J.~Dongarra, ``Practical scalable consensus for pseudo-synchronous
  distributed systems,'' in \emph{SC'15: Proceedings of the International
  Conference for High Performance Computing, Networking, Storage and Analysis},
  2015, pp. 1--12.

\bibitem{bouteiller2015plan}
A.~Bouteiller, G.~Bosilca, and J.~J. Dongarra, ``Plan b: Interruption of
  ongoing mpi operations to support failure recovery,'' in \emph{Proceedings of
  the 22nd European MPI Users' Group Meeting}, 2015, p.~11.

\bibitem{ali2016complex}
M.~M. Ali, P.~E. Strazdins, B.~Harding, and M.~Hegland, ``Complex scientific
  applications made fault-tolerant with the sparse grid combination
  technique,'' \emph{The International Journal of High Performance Computing
  Applications}, vol.~30, no.~3, pp. 335--359, 2016.

\bibitem{Laguna:2014:EUF:2642769.2642775}
\BIBentryALTinterwordspacing
I.~Laguna, D.~F. Richards, T.~Gamblin, M.~Schulz, and B.~R. de~Supinski,
  ``Evaluating user-level fault tolerance for mpi applications,'' in
  \emph{Proceedings of the 21st European MPI Users' Group Meeting}, ser.
  EuroMPI/ASIA '14.\hskip 1em plus 0.5em minus 0.4em\relax New York, NY, USA:
  ACM, 2014, pp. 57:57--57:62. [Online]. Available:
  \url{http://doi.acm.org/10.1145/2642769.2642775}
\BIBentrySTDinterwordspacing

\bibitem{bland2013post}
W.~Bland, A.~Bouteiller, T.~Herault, G.~Bosilca, and J.~Dongarra,
  ``Post-failure recovery of mpi communication capability: Design and
  rationale,'' \emph{The International Journal of High Performance Computing
  Applications}, vol.~27, no.~3, pp. 244--254, 2013.

\bibitem{laguna2016evaluating}
I.~Laguna, D.~F. Richards, T.~Gamblin, M.~Schulz, B.~R. de~Supinski, K.~Mohror,
  and H.~Pritchard, ``Evaluating and extending user-level fault tolerance in
  mpi applications,'' \emph{The International Journal of High Performance
  Computing Applications}, vol.~30, no.~3, pp. 305--319, 2016.

\bibitem{doi:10.1002/cpe.4863}
\BIBentryALTinterwordspacing
S.~Chakraborty, I.~Laguna, M.~Emani, K.~Mohror, D.~K. Panda, M.~Schulz, and
  H.~Subramoni, ``Ereinit: Scalable and efficient fault-tolerance for
  bulk-synchronous mpi applications,'' \emph{Concurrency and Computation:
  Practice and Experience}, vol.~0, no.~0, p. e4863, e4863 cpe.4863. [Online].
  Available: \url{https://onlinelibrary.wiley.com/doi/abs/10.1002/cpe.4863}
\BIBentrySTDinterwordspacing

\bibitem{isc2020reinit}
G.~Georgakoudis, L.~Guo, and I.~Laguna, ``Reinit++: Evaluating the performance
  of global-restart recovery methods for mpi fault tolerance,'' in
  \emph{{ISC}}, 202O.

\bibitem{6114441}
L.~{Bautista-Gomez}, S.~{Tsuboi}, D.~{Komatitsch}, F.~{Cappello},
  N.~{Maruyama}, and S.~{Matsuoka}, ``Fti: High performance fault tolerance
  interface for hybrid systems,'' in \emph{SC '11: Proceedings of 2011
  International Conference for High Performance Computing, Networking, Storage
  and Analysis}, Nov 2011, pp. 1--12.

\bibitem{richards2020quantitative}
D.~Richards, O.~Aaziz, J.~Cook, S.~Moore, D.~Pruitt, and C.~Vaughan,
  ``Quantitative performance assessment of proxy apps and parentsreport for ecp
  proxy app project milestone adcd-504-9,'' Lawrence Livermore National
  Lab.(LLNL), Livermore, CA (United States), Tech. Rep., 2020.

\bibitem{neely2017application}
J.~R. Neely and B.~R. de~Supinski, ``Application modernization at llnl and the
  sierra center of excellence,'' \emph{Computing in Science \& Engineering},
  2017.

\bibitem{bautista2011fti}
L.~Bautista-Gomez, S.~Tsuboi, D.~Komatitsch, F.~Cappello, N.~Maruyama, and
  S.~Matsuoka, ``Fti: high performance fault tolerance interface for hybrid
  systems,'' in \emph{International conference for high performance computing,
  networking, storage and analysis (SC)}, 2011.

\bibitem{doi:10.1177/1094342015623623}
\BIBentryALTinterwordspacing
I.~Laguna, D.~F. Richards, T.~Gamblin, M.~Schulz, B.~R. de~Supinski, K.~Mohror,
  and H.~Pritchard, ``Evaluating and extending user-level fault tolerance in
  mpi applications,'' \emph{The International Journal of High Performance
  Computing Applications}, vol.~30, no.~3, pp. 305--319, 2016. [Online].
  Available: \url{https://doi.org/10.1177/1094342015623623}
\BIBentrySTDinterwordspacing

\bibitem{ispass-13:shao}
Y.~S. Shao and D.~Brooks, ``{{ISA-Independent Workload Characterization and its
  Implications for Specialized Architectures}},'' in \emph{IEEE International
  Symposium on Performance Analysis of Systems and Software (ISPASS)}, 2013.

\bibitem{bland2015lessons}
W.~Bland, H.~Lu, S.~Seo, and P.~Balaji, ``Lessons learned implementing
  user-level failure mitigation in mpich,'' in \emph{2015 15th IEEE/ACM
  International Symposium on Cluster, Cloud and Grid Computing}, 2015.

\bibitem{adam2019checkpoint}
J.~Adam, M.~Kermarquer, J.-B. Besnard, L.~Bautista-Gomez, M.~P{\'e}rache,
  P.~Carribault, J.~Jaeger, A.~D. Malony, and S.~Shende, ``Checkpoint/restart
  approaches for a thread-based mpi runtime,'' \emph{Parallel Computing},
  vol.~85, pp. 204--219, 2019.

\bibitem{shahzad2018craft}
F.~Shahzad, J.~Thies, M.~Kreutzer, T.~Zeiser, G.~Hager, and G.~Wellein,
  ``Craft: A library for easier application-level checkpoint/restart and
  automatic fault tolerance,'' \emph{IEEE Transactions on Parallel and
  Distributed Systems}, vol.~30, no.~3, pp. 501--514, 2018.

\bibitem{doi:10.1177/1094342017711505}
\BIBentryALTinterwordspacing
G.~Bosilca, A.~Bouteiller, A.~Guermouche, T.~Herault, Y.~Robert, P.~Sens, and
  J.~Dongarra, ``A failure detector for hpc platforms,'' \emph{The
  International Journal of High Performance Computing Applications}, vol.~32,
  no.~1, pp. 139--158, 2018. [Online]. Available:
  \url{https://doi.org/10.1177/1094342017711505}
\BIBentrySTDinterwordspacing

\bibitem{6494566}
K.~{Mohror}, A.~{Moody}, G.~{Bronevetsky}, and B.~R. {de Supinski}, ``Detailed
  modeling and evaluation of a scalable multilevel checkpointing system,''
  \emph{IEEE Transactions on Parallel and Distributed Systems}, vol.~25, no.~9,
  pp. 2255--2263, Sep. 2014.

\bibitem{hargrove2006berkeley}
P.~H. Hargrove and J.~C. Duell, ``Berkeley lab checkpoint/restart (blcr) for
  linux clusters,'' in \emph{Journal of Physics: Conference Series}, vol.~46,
  no.~1, 2006, p. 494.

\bibitem{sankaran2005lam}
S.~Sankaran, J.~M. Squyres, B.~Barrett, V.~Sahay, A.~Lumsdaine, J.~Duell,
  P.~Hargrove, and E.~Roman, ``The lam/mpi checkpoint/restart framework:
  System-initiated checkpointing,'' \emph{JHPCA}, vol.~19, no.~4, pp. 479--493,
  2005.

\bibitem{subasi2018unified}
O.~Subasi, T.~Martsinkevich, F.~Zyulkyarov, O.~Unsal, J.~Labarta, and
  F.~Cappello, ``Unified fault-tolerance framework for hybrid task-parallel
  message-passing applications,'' \emph{The International Journal of High
  Performance Computing Applications}, vol.~32, no.~5, pp. 641--657, 2018.

\bibitem{wang2018fault}
Z.~Wang, L.~Gao, Y.~Gu, Y.~Bao, and G.~Yu, ``A fault-tolerant framework for
  asynchronous iterative computations in cloud environments,'' \emph{IEEE
  Transactions on Parallel and Distributed Systems}, vol.~29, no.~8, pp.
  1678--1692, 2018.

\bibitem{cao2016system}
J.~Cao, K.~Arya, R.~Garg, S.~Matott, D.~K. Panda, H.~Subramoni, J.~Vienne, and
  G.~Cooperman, ``System-level scalable checkpoint-restart for petascale
  computing,'' in \emph{2016 IEEE 22nd International Conference on Parallel and
  Distributed Systems (ICPADS)}, 2016.

\bibitem{adam2018transparent}
J.~Adam, J.-B. Besnard, A.~D. Malony, S.~Shende, M.~P{\'e}rache, P.~Carribault,
  and J.~Jaeger, ``Transparent high-speed network checkpoint/restart in mpi,''
  in \emph{Proceedings of the 25th European MPI Users' Group Meeting}, 2018,
  p.~12.

\bibitem{kohl2019scalable}
N.~Kohl, J.~H{\"o}tzer, F.~Schornbaum, M.~Bauer, C.~Godenschwager,
  H.~K{\"o}stler, B.~Nestler, and U.~R{\"u}de, ``A scalable and extensible
  checkpointing scheme for massively parallel simulations,'' \emph{The
  International Journal of High Performance Computing Applications}, vol.~33,
  no.~4, pp. 571--589, 2019.

\bibitem{losada2017resilient}
N.~Losada, I.~Cores, M.~J. Mart{\'\i}n, and P.~Gonz{\'a}lez, ``Resilient mpi
  applications using an application-level checkpointing framework and ulfm,''
  \emph{The Journal of Supercomputing}, vol.~73, no.~1, 2017.

\bibitem{teranishi2014toward}
K.~Teranishi and M.~A. Heroux, ``Toward local failure local recovery resilience
  model using mpi-ulfm,'' in \emph{Proceedings of the 21st european mpi users'
  group meeting}, 2014, p.~51.

\bibitem{katti2018epidemic}
A.~Katti, G.~Di~Fatta, T.~Naughton, and C.~Engelmann, ``Epidemic failure
  detection and consensus for extreme parallelism,'' \emph{The International
  Journal of High Performance Computing Applications}, vol.~32, no.~5, pp.
  729--743, 2018.

\bibitem{Laguna:2014}
\BIBentryALTinterwordspacing
I.~Laguna, D.~F. Richards, T.~Gamblin, M.~Schulz, and B.~R. de~Supinski,
  ``Evaluating user-level fault tolerance for mpi applications,'' in
  \emph{Proceedings of the 21st European MPI Users' Group Meeting}, ser.
  EuroMPI/ASIA '14.\hskip 1em plus 0.5em minus 0.4em\relax New York, NY, USA:
  ACM, 2014, pp. 57:57--57:62. [Online]. Available:
  \url{http://doi.acm.org/10.1145/2642769.2642775}
\BIBentrySTDinterwordspacing

\bibitem{SULTANA20191}
\BIBentryALTinterwordspacing
N.~Sultana, M.~Rüfenacht, A.~Skjellum, I.~Laguna, and K.~Mohror, ``Failure
  recovery for bulk synchronous applications with mpi stages,'' \emph{Parallel
  Computing}, vol.~84, pp. 1 -- 14, 2019. [Online]. Available:
  \url{http://www.sciencedirect.com/science/article/pii/S0167819118303260}
\BIBentrySTDinterwordspacing

\bibitem{Gamell2014Fenix}
\BIBentryALTinterwordspacing
M.~Gamell, D.~S. Katz, H.~Kolla, J.~Chen, S.~Klasky, and M.~Parashar,
  ``Exploring automatic, online failure recovery for scientific applications at
  extreme scales,'' in \emph{Proceedings of the International Conference for
  High Performance Computing, Networking, Storage and Analysis}, ser. SC
  '14.\hskip 1em plus 0.5em minus 0.4em\relax Piscataway, NJ, USA: IEEE Press,
  2014, pp. 895--906. [Online]. Available:
  \url{https://doi.org/10.1109/SC.2014.78}
\BIBentrySTDinterwordspacing

\bibitem{bull2009microbenchmark}
J.~M. Bull, J.~P. Enright, and N.~Ameer, ``A microbenchmark suite for
  mixed-mode openmp/mpi,'' in \emph{International Workshop on OpenMP}.\hskip
  1em plus 0.5em minus 0.4em\relax Springer, 2009.

\bibitem{luszczek2006hpc}
P.~R. Luszczek, D.~H. Bailey, J.~J. Dongarra, J.~Kepner, R.~F. Lucas,
  R.~Rabenseifner, and D.~Takahashi, ``The hpc challenge (hpcc) benchmark
  suite,'' in \emph{Proceedings of the 2006 ACM/IEEE conference on
  Supercomputing}, 2006.

\bibitem{agarwal2014design}
T.~Agarwal and M.~Becchi, ``Design of a hybrid mpi-cuda benchmark suite for
  cpu-gpu clusters,'' in \emph{2014 23rd International Conference on Parallel
  Architecture and Compilation Techniques (PACT)}.\hskip 1em plus 0.5em minus
  0.4em\relax IEEE, 2014.

\bibitem{reussner1998skampi}
R.~Reussner, P.~Sanders, L.~Prechelt, and M.~M{\"u}ller, ``{SKaMPI: A detailed,
  accurate MPI benchmark},'' in \emph{European Parallel Virtual Machine/Message
  Passing Interface Users’ Group Meeting}.\hskip 1em plus 0.5em minus
  0.4em\relax Springer, 1998.

\bibitem{bureddy2012omb}
D.~Bureddy, H.~Wang, A.~Venkatesh, S.~Potluri, and D.~K. Panda, ``{OMB-GPU: a
  micro-benchmark suite for evaluating MPI libraries on GPU clusters},'' in
  \emph{European MPI Users' Group Meeting}.\hskip 1em plus 0.5em minus
  0.4em\relax Springer, 2012.

\end{thebibliography}


\end{document}